\documentclass[acmsmall]{acmart}%
\setcopyright{acmlicensed}% acmcopyright
\usepackage{multirow}%
\usepackage{soul}%
\usepackage{tabularx}%
\usepackage{makecell}%
\usepackage{framed}%
\usepackage{color}%
\usepackage{caption}%
\usepackage{subcaption}%
\newcommand{\NUMTOTALARTICLES}{171}%
\usepackage{fp}% for calculations
\newcommand{\pt}[1]{{%
    \FPeval\resulta{#1/\NUMTOTALARTICLES*100}%
    \FPeval\resulta{round(resulta:1)}%
    \resulta%
}}%
\setcopyright{acmlicensed}%
\copyrightyear{2023}%
\acmJournal{PACMHCI}%
\acmYear{2023}%
\acmVolume{4}%
\acmNumber{CSCW}%
\acmArticle{}%
\begin{document}

\title[Pilot Study Reporting in Crowdsourcing]{%
The State of Pilot Study Reporting in Crowdsourcing:
A Reflection on Best Practices and Guidelines
% A Reflection on Best Practices and Guidelines for Reporting Pilot Studies in Crowdsourcing
}%

\author{Jonas Oppenlaender}%
\email{joppenlu@jyu.fi}%
\orcid{0000-0002-2342-1540}%
\affiliation{%
  \institution{Elisa Corporation}
  \streetaddress{Ratavartijankatu 5}
  \city{Helsinki}
  \country{Finland}
  \postcode{00520}%
}%

\author{Tahir Abbas}%
\email{t.abbas-1@tudelft.nl}%
\orcid{0000-0002-0558-6106}%
\affiliation{%
 \institution{Delft University of Technology}
 \streetaddress{Mekelweg 5}
 \postcode{2628 CD}%
 \city{Delft}
 \country{Netherlands}%
}%

\author{Ujwal Gadiraju}
\email{u.k.gadiraju@tudelft.nl}
\orcid{0000-0002-6189-6539}%
\affiliation{%
 \institution{Delft University of Technology}
 \streetaddress{Mekelweg 5}
 \postcode{2628 CD}%
 \city{Delft}
 \country{Netherlands}%
}%


%%==================================%%
%% abstract %%
%%==================================%%
\begin{abstract}
% Pilot studies are an essential cornerstone of the design of crowdsourcing campaigns.
% Pilot studies are a common method for estimating the rewards from worker's completion times and improving the task design.
% practice to determine crucial factors in crowdsourcing campaigns such as estimating the reward and improving the task design.
Pilot studies are an essential cornerstone of the design of crowdsourcing campaigns, yet they are often only mentioned in passing in the scholarly literature. A lack of details surrounding pilot studies in crowdsourcing research hinders the replication of studies and the reproduction of findings, stalling potential scientific advances.
We conducted a systematic literature review on the current state of pilot study reporting at the intersection of crowdsourcing and HCI research. Our review of ten years of literature included 171~articles published in the proceedings of the Conference on Human Computation and Crowdsourcing (AAAI HCOMP) and the ACM Digital Library.
We found that pilot studies in crowdsourcing research (i.e., \textit{crowd pilot studies}) are often under-reported in the literature.
    Important details, such as the number of workers and rewards to workers, are often not reported.
On the basis of our findings, we reflect on the current state of practice and formulate a set of best practice guidelines for reporting crowd pilot studies in crowdsourcing research.
We also provide implications for the design of crowdsourcing platforms and make practical suggestions for supporting crowd pilot study reporting.
\end{abstract}%
% \ccsdesc[300]{Human-centered computing~Empirical studies in HCI}




\maketitle
%%%%%%%%%%%%%%%%%%%%%%%%%%%%%%%%%%%%%%%%%
%%%%%%%%%%%%%%%%%%%%%%%%%%%%%%%%%%%%%%%%%

% ====================
\section{Introduction}%
% ====================
%
Crowdsourcing is an empirical research area that involves human subjects. The very ingredients that make crowdsourcing a powerful paradigm -- diversity in the background of participating individuals and independence in their opinion \citep{surowiecki2005wisdom} -- also lead to a wide range of behavior and a high variance in performance. It is therefore no surprise that a majority of work in the realms of crowdsourcing research over the last two decades has focused on addressing challenges related to quality \citep{p453-kittur.pdf,kittur2013future,gadiraju2015human,3025453.3026044.pdf}. This well-documented {variability in human behavior and performance while carrying out crowdsourcing tasks} interacts with other task parameters  to shape outcomes, such as the task reward~\citep{yin2015bonus}, task complexity~\citep{yang2016modeling}, task clarity~\citep{p5-gadiraju.pdf}, batch size~\citep{difallah2015dynamics}, and reward schemes~\citep{fan2020crowdco}. Many of such influential configuration parameters of a crowdsourcing campaign are not known before the campaign is launched. Due to this, researchers and practitioners turn to pilot studies to inform their design choices and fine-tune such parameters. Pilot studies are a vital part of crowdsourcing research and researchers often launch one or several small-scale studies before launching the main study.
One typical reason, among others, is to estimate the average completion time of crowdsourced tasks with the aim of appropriately setting the monetary rewards for the larger main study.
In this work, we refer to these small preliminary studies which are often used to calibrate crowdsourcing task design parameters or inform main studies %in one or more ways
as \textbf{crowd pilot studies}.

% ==========================
%%% PROBLEM: MENTIONED ONLY IN PASSING
% ==========================
Despite the important role that crowd pilot studies play in configuring and thereby shaping crowdsourcing studies, from a preliminary review of the literature, we find that it is common for authors to mention crowd pilot studies only in passing.
Crowd pilot studies are often conducted in an ad-hoc manner and details about the crowd pilot studies are seldom reported. %UG will do this: Add a couple of sentences about why this is problematic! (1) replication, reproduction of studies, findings.. fundamental tenets of the scientific method... (2) open science ... (3) defining standard practice as a research community ... transparency)

% ==========================
%%% WHY ARE REPORTING STANDARDS NEEDED?
% ==========================

As a research community that is still evolving~\citep{kittur2013future}, it is important to set the right precedents and establish good practices.
One obstacle to establishing good practices is that crowdsourcing research is conducted in many research communities from different disciplines. The prevailing practices in these different research communities may be highly diverse.
A simple psychological model of why pilot studies are not more often reported in the literature is the availability heuristic \cite{availability}.
    The more researchers report crowd pilot studies in an opaque way, the more frequently other researchers from their community (or other communities) will observe such reporting and read about it in the literature.
Practitioners and researchers in the communities, therefore, may receive signals from the community that the details of the pilot study are not important. A further signal is sent by reviewers and editors
who accept papers with sparse details on pilot studies. 
Therefore, researchers and practitioners may not consider details about pilot study as important or useful for their own work, and may not place importance on reporting their own pilot studies.

% ==========================
% WHY IS THIS A PROBLEM?
% ==========================

% explain why it is problematic or how it affects reproducibility
% the necessity of providing a set of best practices and guidelines for reporting pilot studies in crowdsourcing research.

% ==========================
%%% WHY IS OPAQUE REPORTING A PROBLEM?
% ==========================

This is undesirable for several reasons.
Opaque reporting of pilot studies lies in stark contrast to one of the fundamental tenets of \textit{open science} and recent frameworks such as the Open Science Framework~\citep{foster2017open} -- to make knowledge transparent and accessible~\citep{vicente2018open}.
A lack of transparency on key design parameters of a crowdsourcing campaign hinders future reproduction and replication~\citep{echtler2018open}.
    For instance, readers can glean little from reading that authors `\textit{iterated extensively in pilot studies with crowd workers to strike a balance between simplicity (avoid complex or numerous instructions) and effectiveness (make the layout better)}'~--- a quote from literature reviewed in this work.
    Based on such a description of a crowd pilot study, researchers or practitioners, who may want to learn more about how to achieve a balance between simplicity and effectiveness for their own crowdsourcing study, will arguably be left guessing.
It is worth noting that pilot studies are just as likely to be flawed as any other (main) studies which are expected to withstand the scrutiny of peer-review as a means to ensure quality, reliability, good practice, and a sound scientific method. Such flaws in pilot studies can go unnoticed if they are not reported in sufficient detail.

Another reason why more transparency around pilot studies is warranted is motivated from the crowd workers' perspective.
It is not uncommon for researchers to underestimate the price of a task (we speak from our own experience in this regard). Yet in our literature review, we found it is not very common to make up for these estimation errors (e.g., with bonus payments to workers).
This may contribute to the systematic underpayment of workers \citep{data-driven-analysis-of-workers-earnings-on-amazon-mechanical-turk.pdf,f5caef4f46d2a37d210b9e9811cb207921ba.pdf,wsdmf074-difallahA.pdf}.
Further, crowd pilot studies are very common, yet they are relatively unattractive to many workers as they are small-scale -- that is, done in only small batches of tasks --, and underpaid.
While some workers may be motivated to participate in pilot studies \cite{crowdsourcingcreativity}, other workers may want to avoid them.

Given the prevailing practices and highly interdisciplinary nature of crowdsourcing research, we believe it is not likely for common reporting standards to emerge on their own from within the different research communities, unless the community is alerted about the state of pilot study reporting and incentivized to change their prevailing practices.
%
% such standardization may, one might think, emerge naturally over time.
% For instance, articles in most HCI venues now follow similar structures, and this way of structuring articles has emerged from the community of researchers.
% We argue in the case of crowd pilot studies, reporting standards will \textit{not} naturally emerge in the research community.
% This is because many researchers, use pilot studies to obfuscate design choices that should otherwise be scrutizined in more detail.
%
Our literature review serves this end, by providing guidelines and practical suggestions on how the current state of crowd pilot study reporting could be improved.
We aim toward the development of common reporting standards for crowd pilot studies.
Pilot studies are highly diverse and this, of course, is one reason contributing to the fact that there is no consensus on reporting them in the academic community. Researchers use different ``terms'' to denote pilot studies (i.e., pilot studies or pilot test), and also report them in different sections or in different levels of detail.
Yet in our literature review, we show that there are many commonalities between crowd pilot studies across a multitude of different fields that would allow to develop common reporting standards.
%
% concrete examples of studies in the Introduction section
%
% insights from the literature may.....
%
% Also, they usually report it in very different ways based on different purposes because the goals of pilot studies are
% very diverse (i.e., some are used for validating the feasibility; some are used for iterating their
% system design; some are used for other reasons).
%
%
But while there are guidelines and checklists for running and reporting crowdsourcing studies \citep{2107.13519.pdf,3406865.3418318.pdf,032906PlainLanguageRpt-2.pdf,9948-21517-1-PB.pdf,draws2021checklist}, there is a gap in the  scholarly literature on pilot studies in crowdsourcing research.

In this paper, we aim to address this gap and synthesize the best practices in reporting crowd pilot studies. 
%
%%% METHOD
To this end, we first conducted a systematic literature review.
Our screening of 513 articles downloaded from the ACM Digital Library (ACM-DL) and the proceedings of the AAAI Conference on Human Computation and Crowdsourcing (HCOMP)~-- a premier venue for crowdsourcing related research~-- resulted in a corpus of \NUMTOTALARTICLES~articles.
We systematically analyze this corpus to capture the current state of crowd pilot study reporting in the scholarly literature.
% --------------------
% \subsection{Research Questions}%
\label{sec:researchquestions}%
% --------------------
% UG: Move RQs to the introduction section. It's nicer if we can move the RQs before we enter the Method section where we want to describe how we addressed the RQs.
% Our research investigates how pilot studies are being reported in the scholarly literature.
Our aim is to report and reflect on the current state of pilot study reporting at the intersection of the HCI and crowdsourcing literature.
To this end, we identified whether and to what extent the following information is being reported in articles:
% \begin{framed}
\begin{itemize}
    % \item[RQ1:] Who are the requesters reporting pilot studies?
    % \item[RQ2:] How are crowd pilot studies being referred to in the literature?
    \item[] \textbf{RQ1}: % \hyperref[sec:RQ1]{
        \textit{Why are crowd pilot studies typically conducted?}
        % }
    \item[] \textbf{RQ2}: % \hyperref[sec:RQ2]{
        \textit{How are crowd pilot studies typically reported?}
        %}
    \item[] \textbf{RQ3}: % \hyperref[sec:RQ3]{
        \textit{What do crowd pilot studies report?}
        % }
    % \item[RQ4:] What are the differences in crowd pilot study reporting between research communities?
\end{itemize}
% \end{framed}

While pilot studies are very common in crowdsourcing research, little is known and reported about them in the scholarly literature \citep{3406865.3418318.pdf,2107.13519.pdf}.
Therefore, much of the knowledge from running pilot studies is bound in researchers with experience in crowdsourcing. An experienced researcher may, for instance, decide to not even conduct a crowd pilot study because the researcher's experience will tell what parameters of a crowdsourcing campaign will work best.
We therefore complemented our literature review with a survey study with experienced crowdsourcing researchers to fill the aforementioned gap.
The survey study investigated broader topics not explicitly reported in the scholarly literature:%
% \begin{framed}
\begin{itemize}%
%    \item[RQ5:] What features define a pilot study?
%%%     \item[RQ6:] How often do researchers in academia and industry conduct and report pilot studies, and what are their main goals with regard to pilot studies?
%    \item[RQ6:] What are the factors that promote or obstruct reporting pilot studies?
%    \item[RQ7:] How does the pilot study differ from the main study in terms of worker compensation, cost ratio, platform capabilities, and institution-industry ties?
    % \item reason(s) for conducting the pilot study and reported outcomes,
    % \item whether the number of workers was mentioned and if so, how many,
    % \item whether the pilot study was reported in its own section,
    % \item whether the number of assignments or tasks was mentioned, and if so, how many, and
    % \item the payment to workers participating in the pilot study.
    % \item[RQ4:] \st{What is the motivation for conducting crowd pilot studies?}
    \item[] \textbf{RQ4}: % \hyperref[sec:RQ4]{
        \textit{What makes a ``good'' crowd pilot study?}
        % }
    \item[] \textbf{RQ5}: % \hyperref[sec:RQ5]{
        \textit{What are the factors that promote or obstruct reporting crowd pilot studies?}
        % }
    \item[] \textbf{RQ6}: % \hyperref[sec:RQ6]{
        \textit{How can crowd pilot studies be facilitated with platform-specific features?}
        % }%
\end{itemize}
% \end{framed}

%%% CONTRIBUTION
To the best of our knowledge, our work is the first to provide a detailed investigation on the current state of practice of crowd pilot study reporting in the crowdsourcing and HCI literature. 
Based on the findings of our literature review and survey study, we provide a set of guidelines for reporting crowd pilot studies.
We reflect on the trade-offs around running pilot studies and discuss implications for the design of crowdsourcing platforms. All data % and code
pertaining to our work in this paper are made publicly available for the benefit of the research community and in the spirit of open science.\footnote{\url{https://osf.io/46fxj/?view_only=269ef90006124116ac005392f5389e1a}}

%%% STRUCTURE
Our work is structured as follows.
We first provide a brief review of related literature in Section~\ref{sec:relatedwork}.
We then describe our methodological approach for conducting the systematic literature review and the complementary survey study in Section~\ref{sec:method}.
In Section~\ref{sec:results}, we present the results of our analysis, followed by a reflection and discussion of our findings in Section~\ref{sec:discussion}.
We discuss caveats and limitations of our work in Section~\ref{sec:limitations} and conclude in Section~\ref{sec:conclusion}.

% ====================
\section{Related literature}%
\label{sec:relatedwork}%
% ====================

% --------------------
\subsection{Pilot Studies in Crowdsourcing-based Research}%
% --------------------
%
The crowdsourcing paradigm has seen vast adoption in academia and industry. Crowdsourcing is a cost-effective method for conducting online experiments \citep{jdm10630a.pdf} and user studies \citep{p453-kittur.pdf}.
However, designing an effective crowdsourcing campaign is not an easy task and there are many pitfalls for requesters when designing crowdsourcing campaigns.
For instance, task clarity is one important determinant of work quality~\citep{p5-gadiraju.pdf}.
% Crowdsourcing tasks are subject to the subjective interpretation by the worker.
Many other factors can potentially affect the work quality, such as a task's complexity \citep{p70-Borromeo.pdf}, usability, and accessibility \citep{3411764.3445291.pdf}.

Crowd pilot studies are typically conducted to address these challenges. Crowd pilot studies aim to iteratively design a task and empirically determine design parameters of a crowdsourcing campaign, such as an estimate of the average completion time per task. This estimate can then be used to calculate the price per task for the main study. Before running a pilot study, the average completion time is unknown. Therefore, trial and error is needed to determine an accurate task pricing for microtasks \citep{5283-Article_Text-8384-1-10-20190926.pdf}.
Determining the amount of pay is part of the design of every crowdsourcing campaign involving monetary incentives.

% Besides investigating ways of designing better tasks, one long-running area of crowdsourcing research is the estimation of the optimal price for a crowdsourced task.
% On the one hand, low paid workers may be unreliable \citep{1110.3564.pdf}. On the other hand, it has been found that increasing the pay to workers may not increase the work quality proportionally \citep{1600150.1600175.pdf}.
%     The related problem is referred to as insufficient effort responding \citep{Huang2012_Article_DetectingAndDeterringInsuffici.pdf} or satisficing \citep{10.1177_0013164415627349.pdf}.

Tools and methods have been developed to support requesters in determining the above parameters.
Objective measures like ETA (error-time area) task~\citep{cheng2015measuring} have been proposed
to help researchers accurately structure and price their work.
    ETA empirically models the relationship between time and error rate by manipulating the time that workers have to complete a task~\citep{cheng2015measuring}. The measure proposes that requesters rapidly iterate on task designs and measure whether the changes improve the performance of workers and task outcomes.
% As mentioned in the introduction, error-time-area (ETA)
    % by \cite{cheng2015measuring}
% is an approach to empirically model the relationship between time and error rate~\citep{cheng2015measuring}.
% The ETA measure is derived by manipulating the time that workers have to complete a task.
Requesters may use the ETA measure to rapidly and iteratively test different task designs and measure whether the changes improve the performance of workers and task outcomes.
Besides ETA, other tools supporting requesters in designing tasks and crowdsourcing campaigns have been developed.
\citet{TaskLint} developed a linting tool that automatically uncovers ambiguities in task instructions and supports requesters in writing task instructions with greater clarity. % \citep{TaskLint}.
\citet{3465336.3475109.pdf} proposed methods to computationally assess the clarity of tasks and designed a tool to help requesters improve tasks iteratively. %~\citep{nouri2021iclarify}. 
\citet{3411764.3445382.pdf} presented a system for running and monitoring pilot studies. % \citep{3411764.3445382.pdf}.

However, in practice, the most typical remedy to the above challenges is running informal, small-scale studies with prototypical tasks. % \citep{1707.05645.pdf}.
% Small-scale pilot studies are a standing practice in crowdsourcing research.
These small-scale studies are often conducted iteratively to rapidly uncover issues in the design of the task or to empirically derive estimates of important determinants of the crowdsourcing campaign (such as the task pricing).

% --------------------
\subsection{Guidelines for Conducting and Reporting Crowdsourcing Studies}%
% --------------------
%
Several best practices and guidelines have been developed for requesters to design crowdsourcing campaigns.
These guidelines are motivated with two primary concerns.
    Some authors take the workers' perspective and aim to provide guidelines for requesters to conduct fair and responsible crowd work.
    Other authors provide guidelines from the requester's point of view, aiming to optimize the efficiency, cost, quality, and accuracy of crowdsourced work.

From the requester's perspective,
\citet{beginners_guide} presented a guide and best practices for using crowdsourcing platforms. % ~\citep{beginners_guide}.
These guidelines are primarily meant as a beginner's guide to crowdsourcing.
%%%
\citet{9948-21517-1-PB.pdf} also provided guidelines and examples on using crowdsourcing effectively. % \citep{9948-21517-1-PB.pdf}.
The guidelines take a system development perspective aiming to provide ``design and participation best practices'' guiding the development of crowdsourcing systems.
%%%
\citet{10.1.1.149.6649.pdf} provided a short list of guidelines for designing crowdsourcing studies. The article is scoped to practical aspects when conducting relevance evaluations. % \citep{10.1.1.149.6649.pdf}.
%%%
\citet{032906PlainLanguageRpt-2.pdf} presented guidelines for writing clear instructions for voters and poll workers. % ~\citep{032906PlainLanguageRpt-2.pdf}.
While this report is not written for the crowdsourcing domain, the report provides takeaways for writing clear instructions to crowd workers. \citet{gadiraju2015understanding} explore the different ways in which tasks can be exploited by unreliable workers in surveys and propose task design guidelines to thwart such behavior and ensure quality control. \citet{5283-Article_Text-8384-1-10-20190926.pdf} introduced a means to help requesters in automatically paying workers a minimum wage by adding a one-line script tag to their task HTML on Amazon Mechanical Turk (MTurk). \citet{draws2021checklist} proposed a checklist as a practical tool that requesters can use to improve their task designs by mitigating cognitive biases of workers and appropriately describe potential limitations of collected data.

Guidelines written with the workers' perspective in mind are fewer in number.
%%%
For instance, Dynamo by \citet{DynamoCHI2015.pdf}
% took the workers' perspective and
provided worker-generated ``Guidelines for Academic Requesters'' for ethical research on Amazon Mechanical Turk \citep{Guidelines_for_Academic_Requesters.pdf}.
The guidelines
% hosted by Northwestern University's Institutional Review Board (IRB) Office
aim to provide guidance for requesters on ``how to be a good requester,'' fair payment, and other aspects of fair crowd work.
%%%
\citet{doi:10.1080/08956308.2017.1325689} formulated key principles for effective communication with workers in crowdsourcing contests. % ~\citep{doi:10.1080/08956308.2017.1325689}.
%%%
In a similar vein, the ``ground rules'' hosted at \citeauthor{ground_rules} aim to provide guidance ``for a prosperous and fair cooperation between crowdsourcing companies and crowdworkers'' \citep{ground_rules}.
%%%
%%%
Besides the above documents, guidance and feedback for requesters can also be found on worker-focused websites and in online forums, such as Turkopticon~\citep{turkopticon.pdf} and Turker Nation~\citep{f5caef4f46d2a37d210b9e9811cb207921ba.pdf}.

% http://faircrowd.work
% Fair pay is, however, only one factor affecting the workers' precarious work situation.
% Professional crowd workers spend a considerable time of their work day on searching for tasks \citep{3476060.pdf}. This labor is called hidden, because it does not manifest on the platform and is unpaid.
% Another source of hidden labor is the qualification work that workers need to undertake to qualify for more tasks \citep{2105.12762.pdf}.
% % hidden work
% % ghost work
% In order to avoid unfair qualification labor on crowdsourcing platforms \citep{2105.12762.pdf},

However, when it comes to reporting crowdsourcing studies, little guidance is available in the scholarly literature \citep{3406865.3418318.pdf,2107.13519.pdf}.
To the best of our knowledge, there are only two papers providing guidance on how to report crowdsourcing studies and experiments.
% Both were published only recently. 
%
\citet{3406865.3418318.pdf} proposed DREC, a datasheet for reporting experiments in crowdsourcing. % ~\citep{3406865.3418318.pdf}.
Based on an analysis of a sample of 15~scientific articles, 
the authors provided a glimpse into the state of reporting on crowdsourcing studies. The authors found that details of crowdsourcing studies are often not being reported in scientific articles. %have to often be inferred.
The authors created a taxonomy of attributes relevant to crowdsourcing studies aiming to support requesters in reporting crowdsourcing experiments.
\citeauthor{2107.13519.pdf} later extended the DREC taxonomy in scope and provided a checklist for reporting crowdsourcing experiments, based on an analysis of the literature % 171 crowdsourcing experiments
\citep{2107.13519.pdf}.
The article examines the state of reporting on crowdsourcing experiments and offers guidance for requesters.

In relation to these two studies, there is an overlap with our work in that the authors make recommendations for the reporting of key statistics of a crowdsourcing campaign, such as the number of participants.
However, the checklist is clearly scoped to reporting the results of the main crowdsourcing experiment.
The section on pilot studies (Item No. 6) in \citet{2107.13519.pdf} is very short and only recommends to ``[d]escribe if pilot studies were performed before the main experiment'' (p. 30). %~\citep{2107.13519.pdf}.

Our work connects to the above two guidelines by providing an extensive and in-depth investigation on the state and practice of pilot study reporting as well as detailed recommendations for reporting crowd pilot studies.%
%
%
%
%
% ====================
\section{Method}%
\label{sec:method}%
% ====================
We conducted a systematic literature review~\citep{Petticrew_Roberts} to investigate how pilot studies are being reported in the HCI and crowdsourcing literature (%cf.
sections \ref{sec:scope}--\ref{sec:analysis}).
The literature review is complemented with an online survey with requesters (see Section~\ref{sec:survey}).%
%
%
% --------------------
\subsection{Scope of the Literature Review}%
\label{sec:scope}%
% --------------------
Before we started our literature review, we needed to clearly define the research questions (see Section \ref{sec:researchquestions}) and delineate the boundaries of the review \citep{10.1002/9780470754887.ch2}.
% To this end, we define `crowd pilot study' in this section to provide clarity to our subject of study.
Our work focuses specifically on pilot studies that are crowdsourced to a group of diverse and independent individuals online.
Throughout the remainder of this work, we refer to this type of pilot studies as \textbf{crowd pilot studies}.
This concept has two components (`pilot study' and `crowd') which we clarify and define in the following two sections.%
\subsubsection{Pilot study}%
From the onset of our literature review, we were particularly interested in small-scale formative pilot studies in the crowdsourcing domain. The way these studies are being reported in the scholarly literature is often opaque and we wanted to illuminate researchers' practices around reporting pilot studies. This is important % due to ...
because opaque reporting of pilot studies may obfuscate information and -- from a systemic perspective -- attenuate the spread of best practices.
The first iteration of our literature review -- as a scoping review \citep{10.1002/9780470754887.ch2} -- found over 100 articles reporting small-scale pilot studies in a formative way. However, in this scoping review, we found that a considerable amount of pilot studies are also being conducted for summative purposes. We therefore % decided to
extended the scope of our literature review to a broader definition of pilot studies that better captures the state of reporting on both formative and summative studies.

Many different user studies and experiments have been conducted on crowdsourcing platforms.
In our work, a pilot study is any small-scale or larger scale experiment or study % , or prototype
that is being conducted to inform the design of a prototype, validate a proof of concept, or for other formative or summative reasons.
The scope of our work is defined by the authors' use of the term crowdsourcing in combination with the `pilot' keyword.
    For instance, user studies on crowdsourcing platforms are only included in our literature corpus if the studies were referred to by the authors as pilot studies.%
\subsubsection{Crowd}%
\label{sec:crowd}%
The pilot studies could be conducted internally by the researchers within a lab setting or with an external crowd~\citep{gadiraju2017crowdsourcing}. In our work, we exclusively focus on pilot studies conducted with an external crowd. This external crowd needs to consist of people other than the researchers (otherwise it would be considered an ``internal pilot study'').
%  umbrella term
Crowdsourcing comes in many different forms (e.g., crowdfunding, contests, microtasking, among others). 
Our work follows a broad and integrative definition of the crowd. 
Our literature review includes studies conducted on traditional paid microtasking platforms, situated and mobile crowdsourcing \citep{GONCALVES20171,10.1145/2642918.2647362}, or on other paid and unpaid crowdsourcing platforms with different types of participants (e.g., students and volunteers).%
% The crowd does not necessarily need to receive tasks online, since mobile and situated crowdsourcing \citep{GONCALVES20171,10.1145/2642918.2647362} are also types of crowdsourcing that we consider in our research.
%
%
%
% --------------------
\subsection{Creating a Corpus of Relevant Articles}%
\label{sec:creatingcorpus}%
% --------------------
% \subsubsection{Bibliographic Databases}
We limited our search to articles published in the past ten years (2012 -- 2021).
    The time frame of ten years was chosen to provide a representative window into current best practices that have emerged since the inception of crowdsourcing.
    % Second reason for limiting the time frame: HCOMP is conference since 2012 (prior that in workshop format). The farther back, the more articles cannot be found.

Our search was conducted in two bibliographic sources. First, we downloaded all articles (excluding posters) from the Proceedings of the Conference on Human Computation and Crowdsourcing (HCOMP), widely considered as a primary venue for crowdsourcing related research.
This resulted in 215 articles (of which two articles from the oldest proceedings could not be retrieved).
Second, we searched the Digital Library of the Association for Computing Machinery (ACM).
    The ACM-DL is the document storage for all articles published by the ACM and therefore covers a wide range of different conferences and journals (including the ACM CHI and ACM CSCW conferences where premier works on crowdsourcing research at the intersection of HCI are commonly published). 
The search in the ACM-DL used the following query string:

% 288 Results for: [Abstract: crowdsourc*] AND [Full Text: pilot] AND [Publication Date: (01/01/2013 TO 12/31/2021)]
% +
% 12 Results for: [Abstract: crowdsourc*] AND [Full Text: pilot] AND [Publication Date: (01/01/2012 TO 12/31/2012)]
% \begin{framed}%
\begin{quote}%
% \begin{itemize}%
    \textit{query:} \textit{Fulltext:(pilot) AND Abstract:(crowdsourc*)}
    \\
    \textit{filter:} \textit{Article Type: Research Article, \\
        Publication Date (01/01/2012 to 12/31/2021)}%
        % ACM Content: DL
% \end{itemize}%
    % \textit{[Abstract: crowdsourc*] AND [Full Text: pilot] AND [Publication Date: (01/01/2012 TO 12/31/2021)]}
\end{quote}%
% \end{framed}
The choice of the `pilot' keyword reflects the scope of our literature review.
We found that limiting our search to occurrences of ``crowdsourcing'' (or derivations thereof) in the abstract was a good compromise between identifying relevant literature and avoiding false positives (e.g., studies that mention crowdsourcing in the related work or the references).

Our aim was to arrive at a representative coverage of articles in the literature~\citep{giants.pdf}.
We focused on the proceedings of AAAI HCOMP and the ACM Digital Library due to their prominence and influence in the crowdsourcing research community. These venues are known for their rigorous peer-review processes and attract a wide range of high-quality submissions from researchers globally, making them representative sources for our analysis. While we acknowledge that relevant literature might also exist in the IEEE Xplore library and other repositories, our choice was driven by the desire to capture the core developments and trends in crowdsourcing from these leading communities. % Future studies could consider a more expansive scope to include other significant repositories.
Our search resulted in a corpus of 513 articles (213 articles at HCOMP and 300 articles in the ACM-DL).

% --------------------
\subsection{Article filtering and exclusion criteria}%
\begin{figure}[!h]%
\centering%
\includegraphics[width=\linewidth]{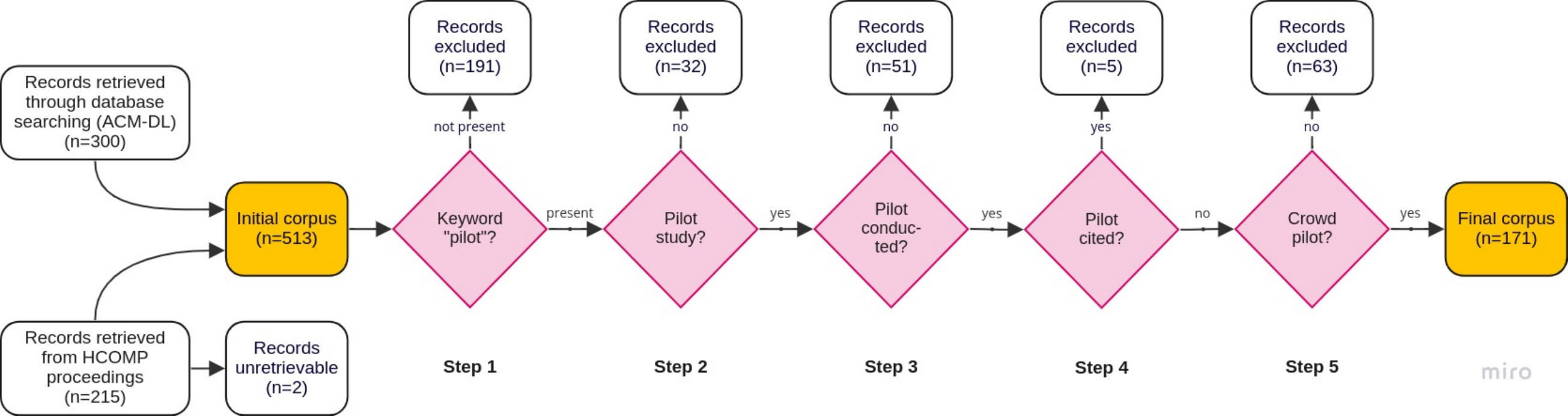}%
\caption{Corpus screening procedure.}%
% \Description{Flowchart visualizing the corpus screening procedure in five consecutive steps.}
\label{fig:PRISMA}%
\end{figure}

% --------------------
% We provide details on a number of decisions made in our analysis of the 514 articles.
We filtered the corpus of 513 articles in five consecutive steps.
The steps are depicted in the flowchart in \autoref{fig:PRISMA}.
% -----
% -----
% ----------------
% \input{figures/FIG-PRISMA}%
% ----------------

% Step 1
% \subsubsection{Step 1: Identifying pilot studies}
% In a first step, we searched for the keyword ``pilot'' in each article.
First, we identified whether the keyword ``pilot'' was present in the article. Articles from the HCOMP proceedings that did not include this term were excluded ($n=191$). This step also excluded articles from the ACM-DL where the pilot keyword was only mentioned in the references.
%
% Step 2
Second, we identified whether the term ``pilot'' refers to a pilot study or experiment, or whether it denoted something else (e.g., `Palm Pilot' or `airplane co-pilot').
% This step also excluded, for instance, pilot surveys.
This step excluded 32~articles.
%
% Step 3
Third, we identified whether the pilot study was conducted by the authors in the article. This step excluded articles that mentioned pilot studies in the related work section or only provided recommendations for conducting pilot studies, without conducting one in the article. Articles that did not conduct a pilot study were excluded ($n=51$).
%
% Step 4
Next, we excluded articles in which authors referenced and discussed their pilot study in other articles. We decided not to conduct a backward reference search because the cited articles did not contain our search keyword and we wanted to focus on full articles. This step excluded five articles.
%
% Step 5
Finally, we identified whether the pilot study involved crowd workers.
As mentioned in Section~\ref{sec:crowd}, we apply a broad definition of crowd work that includes everything from situated crowdsourcing, citizen science (e.g., Zooniverse), volunteering, to paid microwork.
We specifically focused on pilot studies conducted with a crowd, excluding other participants (such as experts) in our analysis.
If the pilot study did not include a pilot study conducted with a crowd, the article was excluded ($n=63$).
If it was not fully clear from the authors' statements whether the pilot study involved crowdsourcing, we included the article in our analysis ($n=2$).
% 2858036.2858279.pdf
% 3321700.pdf = early pilots
    % For instance, \citet{3025453.3025922.pdf} report that three \textit{``experiments and pilots''} in the context of a study on Amazon Mechanical Turk were conducted, but the authors later used the terms ``internal piloting'' and ``piloting'' which does not allow to determine whether the pilot involved crowd workers \citep{3025453.3025922.pdf}.

We validated the robustness of our 5-step literature screening procedure by conducting a sensitivity analysis.
The sensitivity analysis involved slightly altering the criteria for the first step and then applying the full 5-step filtering process on a subset of articles. More specifically, we expanded the first step to include ``preliminary,'' ``initial,'' and ``formative'' as search keywords, besides ``pilot,'' and observed the impact on the sampled articles.
As subset for the sensitivity analysis, we selected two HCOMP proceedings (2017 and 2018) with 46 articles (8.97\% of the initial corpus).
Validation of our filtering methodology showed strong internal consistency.
The sensitivity analysis revealed that modifications in the search keyword resulted in less than a 5\% change in the subset of papers (2 out of 48; 4.35\%), confirming the robustness of our filtering process.
The sensitivity analysis also confirmed our initial suspicion that ``pilot'' is the most commonly used term to refer to a crowd pilot study.

Our final set of literature comprises \NUMTOTALARTICLES~articles (23~articles from HCOMP and 148~articles from the ACM-DL).
Throughout our work, direct quotes from articles are printed in \textit{italics}.
The literature corpus was analyzed as follows.%
%
%
%
% --------------------
\subsection{Analyzing the Corpus}%
\label{sec:analysis}%
% --------------------
We started our analysis by familiarizing ourselves with the articles. To this end, we manually extracted all verbatim statements that mentioned the pilot keyword from each article together with the surrounding context. Typically, there were only few mentions of the pilot keyword in one paragraph or short sentences of the article.
If there were pilot studies with multiple participant samples, we focused on the pilot studies conducted with the crowd in our analysis.

For answering our research questions, we followed an inductive approach based on grounded theory \citep{groundedtheory}.
We iteratively revisited the verbatim statements to identify what could be reported about the articles (e.g., the number of pilot studies conducted in the article or whether payment to workers was reported). If our research questions could not be answered from the verbatim statements, we revisited the article for closer reading.

Coding was straightforward in cases when variables were binary (e.g., deciding whether the pilot study was reported in its own section) or when information had to be extracted (e.g., the year of publication). This straightforward coding required only one coder and no inter-rater agreement was calculated~\citep{McDonald_Reliability_CSCW19.pdf}.
Other cases were more challenging. These cases were analyzed by two post-doctoral researchers.
The coding was developed iteratively and from the bottom-up in several coding passes.
The first coding iteration stayed close to the information provided in the articles.
This first iteration allowed us to form an understanding of categories in the data, which we then iteratively grouped into codes.
The coding results were frequently discussed among all authors which resulted in codes being adjusted and articles iteratively being re-coded.
The coding was done in an Excel sheet which was then used to produce graphs in R. All data and code relevant to this process will be shared publicly for the benefit of further research, in the spirit of Open Science.\footnote{\url{https://osf.io/46fxj/?view_only=269ef90006124116ac005392f5389e1a}.}

% While the majority of the information was coded using open coding, closed coding was also used (e.g., to decide on the type of sections in which authors report the pilot study.
    % For grouping conference venues and journals, we adopted subject terms from Worldcat.
% In some cases, the analysis involved interpretation and information had to be inferred. For instance, it was often not mentioned what crowdsourcing platform was used in the pilot study. If the crowdsourcing platform was mentioned in the main study of the article, we used this platform.%
%
%
%
% --------------------
\subsection{Survey Study}%
\label{sec:survey}%
% --------------------
%
{%
It is noteworthy that from the literature alone, we cannot discern anything about authors who decide not to report pilot studies.
}
Therefore, our literature review can only provide incomplete insights into the prevalence of pilot studies and the requesters' motivations for running pilot studies. To limit if not alleviate this publication bias~\citep{Petticrew_Roberts.Ch7} of our literature review, we complemented our analysis with an online survey study with crowdsourcing researchers in academia and industry.

The survey was implemented on Qualtrics. Participants' consent was collected before starting the survey.
Participation was incentivized with a raffle of 10 Amazon gift cards (each worth US\$15).
The survey included 25 items and was estimated to take between 10 and 15 minutes. Many questions were closed-ended (with an option to enter an open-ended response, if preferred or needed) to be mindful of researchers' time and not overburden the participants. The open-ended survey items focused on two key areas: 1) the motivation for running pilot studies and 2) the requester's practices around reporting pilot studies.
The former included questions about the participants' motivation for conducting pilot studies, what they consider as a ``good'' pilot study, and criteria for running pilot studies.
The latter asked what factors promote or obstruct the pilot study reporting and possible features of a crowdsourcing platform that could support requesters in running pilot studies.
Participants with experience in collaborating with industry were asked about differences between pilot studies in industry and academia.
% We also collected demographics about the participants (e.g., research area and experience).

We aimed to invite researchers and industry professionals with experience in crowdsourcing.
For this reason, we followed a mix of snowball \citep{euclid.aoms.1177705148.pdf} and convenience sampling \citep{convenience-sampling} to disseminate the survey study to experienced researchers in academia and industry.
We also announced the study in communities dedicated to human computation and crowdsourcing, including the HCOMP Slack Community %\footnote{https://hcomp.slack.com},
(with 396 members at the time of writing) and the Google Group on Crowdsourcing and Human Computation (with 579 members).
The survey had valid responses from twelve participants, but we excluded one because she did not consent to the study.  
Participants included researchers from academia and industry with a background in computer science, human-computer interaction, and design.
The sample includes five Ph.D. students, one postdoctoral researcher, and researchers at the professor-level.
Participants had between 2 and over 12 years of experience in crowdsourcing research.

% ====================
\section{The state of crowd pilot study reporting}
    % in HCI and Crowdsourcing Research}%
\label{sec:results}%
% ====================
%
In this section, we first provide an overview of the literature corpus before we turn to answering our research questions in the subsequent sections.

% --------------------
\subsection{Literature Corpus}%
\begin{figure}[!h]%
\centering%
% \begin{subfigure}[b]{1\textwidth}%
% \centering%
\includegraphics[width=.75\linewidth]{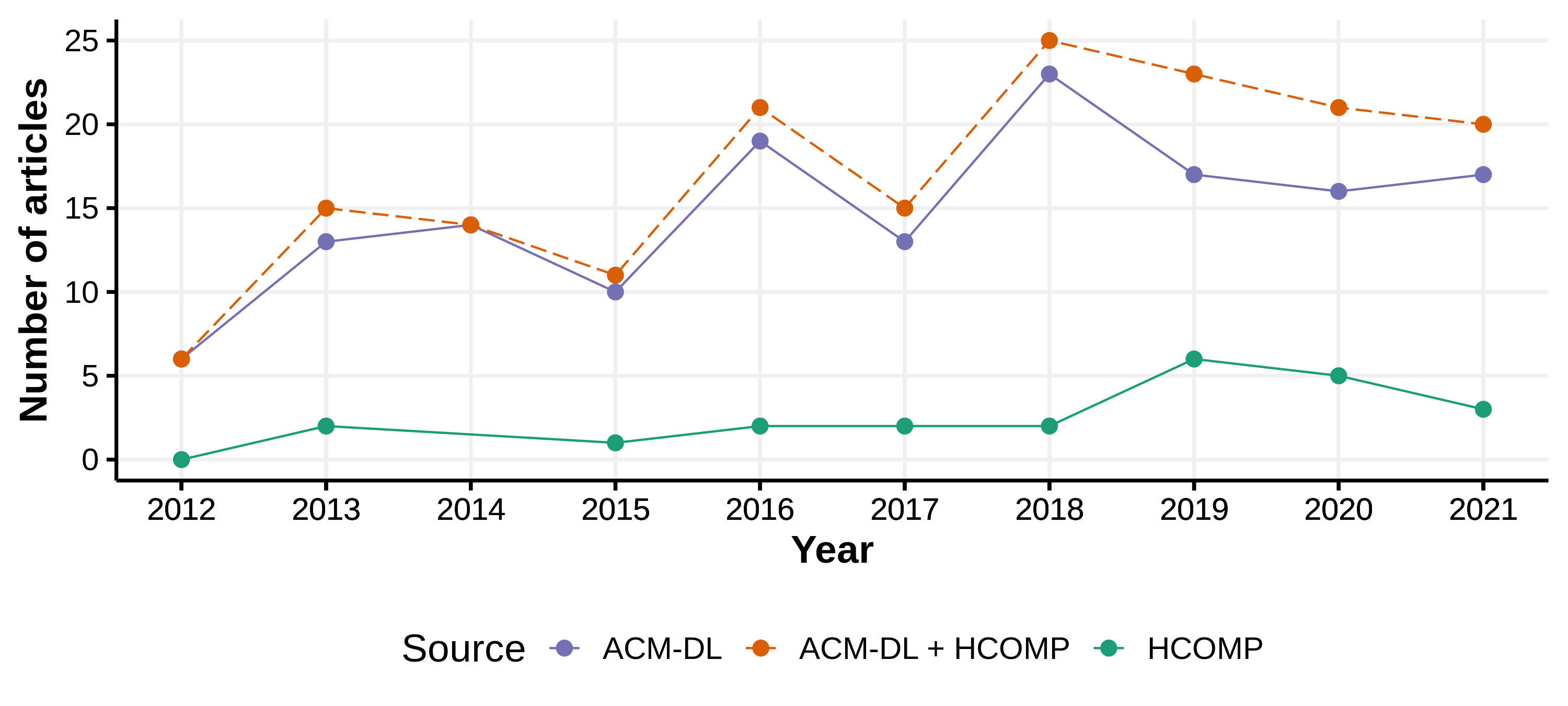}%
% \subcaption{A subfigure}\label{fig:a}
\caption{Articles reporting crowd pilot studies in the ACM Digital Library and HCOMP proceedings.}%
    % \Description{Articles reporting crowd pilot studies in the ACM Digital Library and HCOMP proceedings.}
\label{fig:years}%
% \end{subfigure}%
% \hfill%
% \begin{subfigure}[b]{1\textwidth}%
% \centering%
% \includegraphics[width=.75\linewidth]{figures/images/pNumPages2.png}%
% %             \subcaption{Another subfigure}\label{fig:b}
% \subcaption{Number of pages in articles reporting crowd pilot studies.\\}%
%     % \Description{Number of pages in articles reporting crowd pilot studies.}
% \label{fig:pages}%
%         \end{subfigure}%
% \caption{Overview of articles per year and number of pages within the articles.}%
\label{fig:1}%
\end{figure}

\label{sec:corpus}%
% --------------------
The literature corpus includes \NUMTOTALARTICLES~articles (see \autoref{tab:publications}).
%
% \autoref{fig:years} depicts the number of crowd pilot studies reported per year.
We find the number of articles reporting crowd pilot studies has more than tripled in the past decade (see \autoref{fig:years}).

% -----
%
% \begin{table*}[h]%
% \centering%
% \caption{Research articles reporting crowd pilot studies per year.}%
% \label{tab:publications}%
% \small%
% \begin{tabularx}{.95\textwidth}{lX}%
% \toprule%
%     Year & Articles \\
% \midrule%
%     2012 & \input{years/2012} \\
%     2013 & \input{years/2013} \\
%     2014 & \input{years/2014} \\
%     2015 & \input{years/2015} \\
%     2016 & \input{years/2016} \\
%     2017 & \input{years/2017} \\
%     2018 & \input{years/2018} \\
%     2019 & \input{years/2019} \\
%     2020 & \input{years/2020} \\
%     2021 & \input{years/2021} \\
% \bottomrule%
% \end{tabularx}%
% \end{table*}%
%

\begin{table}[h]%
% \sidewaystablefn%
\begin{center}%
% \begin{minipage}{\textheight}
\caption{Research articles reporting crowd pilot studies per year.}\label{tab:publications}%
\begin{tabularx}{\textwidth}{@{\extracolsep{\fill}}lX@{\extracolsep{\fill}}}%
\toprule%
    Year & Articles \\
\midrule%
    \scriptsize 2012 & \scriptsize \cite{2384916.2384941.pdf},
\cite{2348283.2348400.pdf},
\cite{2145204.2145310.pdf},
\cite{2442576.2442591.pdf},
\cite{2145204.2145354.pdf},
\cite{2384616.2384624.pdf} \\
    \scriptsize 2013 & \scriptsize \cite{2528394.2528396.pdf},
\cite{13072-Article_Text-16589-1-2-20201228.pdf},
\cite{2461912.2462002.pdf},
\cite{2483669.2483676.pdf},
\cite{2470654.2470685.pdf},
\cite{2470654.2470686.pdf},
\cite{2493432.2493481.pdf},
\cite{2513383.2513448.pdf},
\cite{2502081.2502221.pdf},
\cite{2470654.2470743.pdf},
\cite{13075-Article_Text-16592-1-2-20201228.pdf},
\cite{2504776.2504796.pdf},
\cite{2505515.2505695.pdf},
\cite{2464464.2464482.pdf},
\cite{2517899.2517930.pdf} \\
    \scriptsize 2014 & \scriptsize \cite{2594776.2594788.pdf},
\cite{2598510.2598514.pdf},
\cite{2642918.2647403.pdf},
\cite{2669485.2669511.pdf},
\cite{2598510.2598587.pdf},
\cite{2596695.2596713.pdf},
\cite{2556288.2556986.pdf},
\cite{2661829.2661918.pdf},
\cite{2555691.pdf},
\cite{2660114.2660126.pdf},
\cite{2565585.2565608.pdf},
\cite{2660114.2660124.pdf},
\cite{2541288.pdf},
\cite{2600428.2609577.pdf} \\
    \scriptsize 2015 & \scriptsize \cite{2756406.2757741.pdf},
\cite{2740908.2745401.pdf},
\cite{2717513.pdf},
\cite{2736277.2741102.pdf},
\cite{2787622.2787709.pdf},
\cite{2766929.pdf},
\cite{13239-Article_Text-16756-1-2-20201228.pdf},
\cite{2702123.2702608.pdf},
\cite{2772879.2772947.pdf},
\cite{2786451.2786492.pdf},
\cite{2810188.2810190.pdf} \\
    \scriptsize 2016 & \scriptsize \cite{2998476.2998498.pdf},
\cite{2993148.2993200.pdf},
\cite{2897366.pdf},
\cite{2858036.2858198.pdf},
\cite{2897370.pdf},
\cite{2858036.2858327.pdf},
\cite{2858036.2858144.pdf},
\cite{13273-Article_Text-16790-1-2-20201228.pdf},
\cite{2993148.2993190.pdf},
\cite{2818048.2819979.pdf},
\cite{13287-Article_Text-16804-1-2-20201228.pdf},
\cite{2858036.2858321.pdf},
\cite{2903138.pdf},
\cite{2858036.2858279.pdf},
\cite{2984511.2984578.pdf},
\cite{2858036.2858374.pdf},
\cite{2858036.2858572.pdf},
\cite{2872427.2883035.pdf},
\cite{2971485.2971492.pdf},
\cite{2872427.2883036.pdf},
\cite{2818048.2835201.pdf} \\
    \scriptsize 2017 & \scriptsize \cite{3134664.pdf},
\cite{3025453.3026044.pdf},
\cite{3025453.3025922.pdf},
\cite{3025453.3025870.pdf},
\cite{3130914.pdf},
\cite{3130931.pdf},
\cite{13313-Article_Text-16830-1-2-20201228.pdf},
\cite{2998181.2998311.pdf},
\cite{13296-Article_Text-16813-1-2-20201228.pdf},
\cite{3021460.3021483.pdf},
\cite{2998181.2998332.pdf},
\cite{3126594.3126648.pdf},
\cite{3134744.pdf},
\cite{2998181.2998234.pdf},
\cite{CSI-SE.2017.2.pdf} \\
    \scriptsize 2018 & \scriptsize \cite{3209542.3209558.pdf},
\cite{3173574.3173884.pdf},
\cite{3209281.3209347.pdf},
\cite{3209624.pdf},
\cite{3242587.3242598.pdf},
\cite{3230670.pdf},
\cite{3173574.3174216.pdf},
\cite{3274311.pdf},
\cite{3176349.3176381.pdf},
\cite{3173574.3174141.pdf},
\cite{3173574.3173850.pdf},
\cite{3269206.3271779.pdf},
\cite{3287050.pdf},
\cite{13322-Article_Text-16839-1-2-20201228.pdf},
\cite{3284432.3284469.pdf},
\cite{3238147.3240727.pdf},
\cite{3173574.3173846.pdf},
\cite{13325-Article_Text-16842-1-2-20201228.pdf},
\cite{3274423.pdf},
\cite{3173574.3173806.pdf},
\cite{3274428.pdf},
\cite{3178876.3186031.pdf},
\cite{3274447.pdf},
\cite{3230665.pdf},
\cite{3209978.3210064.pdf} \\
    \scriptsize 2019 & \scriptsize \cite{3328906.pdf},
\cite{3359227.pdf},
\cite{3290605.3300482.pdf},
\cite{3321700.pdf},
\cite{3349611.3355545.pdf},
\cite{3290605.3300892.pdf},
\cite{3290605.3300742.pdf},
\cite{3359238.pdf},
\cite{5272-Article_Text-8373-2-10-20191003.pdf},
\cite{3290605.3300422.pdf},
\cite{3365610.3365621.pdf},
\cite{5281-Article_Text-8382-1-10-20190926.pdf},
\cite{5264-Article_Text-8365-1-10-20190926.pdf},
\cite{5268-Article_Text-8369-1-10-20190926.pdf},
\cite{3357384.3357976.pdf},
\cite{3290605.3300750.pdf},
\cite{5271-Article_Text-8372-1-10-20190926.pdf},
\cite{3290605.3300899.pdf},
\cite{3301414.pdf},
\cite{3306618.3314270.pdf},
\cite{3308558.3313569.pdf},
\cite{5283-Article_Text-8384-1-10-20190926.pdf},
\cite{3359158.pdf} \\
    \scriptsize 2020 & \scriptsize \cite{7459-Article_Text-10896-1-10-20200930.pdf},
\cite{3434175.pdf},
\cite{3313831.3376789.pdf},
\cite{3415181.pdf},
\cite{3366423.3380063.pdf},
\cite{3313831.3376365.pdf},
\cite{3340531.3412782.pdf},
\cite{3418196.pdf},
\cite{3403931.pdf},
\cite{3313831.3376799.pdf},
\cite{3394978.pdf},
\cite{3340531.3411863.pdf},
\cite{3366424.3383546.pdf},
\cite{7470-Article_Text-10869-1-10-20200930.pdf},
\cite{3397271.3401112.pdf},
\cite{7471-Article_Text-10867-1-10-20200930.pdf},
\cite{3415176.pdf},
\cite{7472-Article_Text-10866-1-10-20200930.pdf},
\cite{3399715.3399827.pdf},
\cite{3375187.pdf},
\cite{7479-Article_Text-10849-1-10-20200925.pdf} \\
    \scriptsize 2021 & \scriptsize \cite{3484828.pdf},
\cite{3451161.pdf},
\cite{3479550.pdf},
\cite{18939-Article_Text-22705-1-2-20211004.pdf},
\cite{3411764.3445493.pdf},
\cite{3479572.pdf},
\cite{3503516.3503533.pdf},
\cite{3411764.3445344.pdf},
\cite{3411764.3445507.pdf},
\cite{3411764.3445637.pdf},
\cite{3480965.pdf},
\cite{3465336.3475109.pdf},
\cite{3479586.pdf},
\cite{3437963.3441831.pdf},
\cite{3411764.3445782.pdf},
\cite{3411764.3445557.pdf},
\cite{3411764.3445399.pdf},
\cite{3474381.pdf},
\cite{18947-Article_Text-22713-1-2-20211004.pdf},
\cite{18948-Article_Text-22714-1-2-20211004.pdf} \\
\bottomrule%
\end{tabularx}%
% \footnotetext{Note: This is an example of table footnote this is an example of table footnote this is an example of table footnote this is an example of~table footnote this is an example of table footnote.}
% \footnotetext[1]{This is an example of table footnote.}
% \end{minipage}
\end{center}%
% \end{sidewaystable}
\end{table}%

%
% -----

% -----
% -----

\begin{figure}[!h]%
\centering%
\includegraphics[width=1\linewidth]{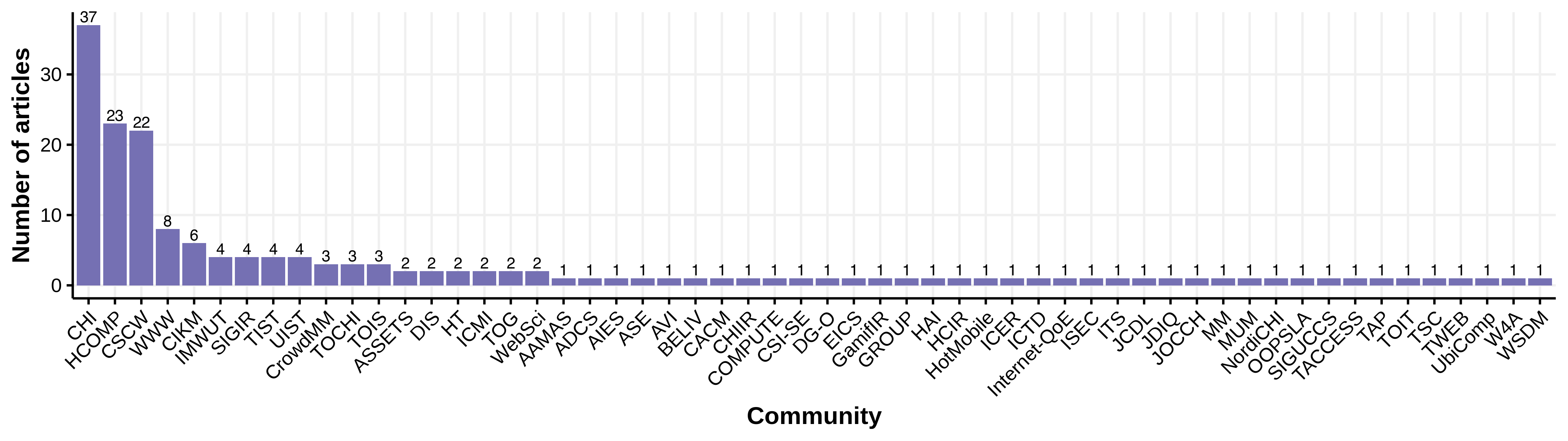}%
\caption{Conference venues and journals in the corpus.}
    % \Description{Conference venues and journals in the corpus.}
\label{fig:communities}%
\end{figure}

The corpus includes articles published in academic conferences ($n=144$), journals ($n=24$), and workshops ($n=3$).
The articles have between 2 and 41~pages (including references and appendices). % ; see \autoref{fig:pages}
The distribution of articles over the different venues is long-tailed (see \autoref{fig:communities}).
About half of the articles ($n=82$, \pt{82}\%) are published in the three conferences on Human Factors in Computing Systems (CHI), Human Computation and Crowdsourcing (HCOMP), and Computer Supported Cooperative Work (CSCW), with other conference venues and journals reporting significantly fewer pilot studies,
${\chi}^2(70, N=171)=923.06$, $p<0.000$.
% X-squared = 923.06, df = 70, p-value < 2.2e-16
% (see \autoref{fig:communities}).
% -----
% -----

\begin{figure}[!h]%
\centering%
\includegraphics[width=.85\linewidth]{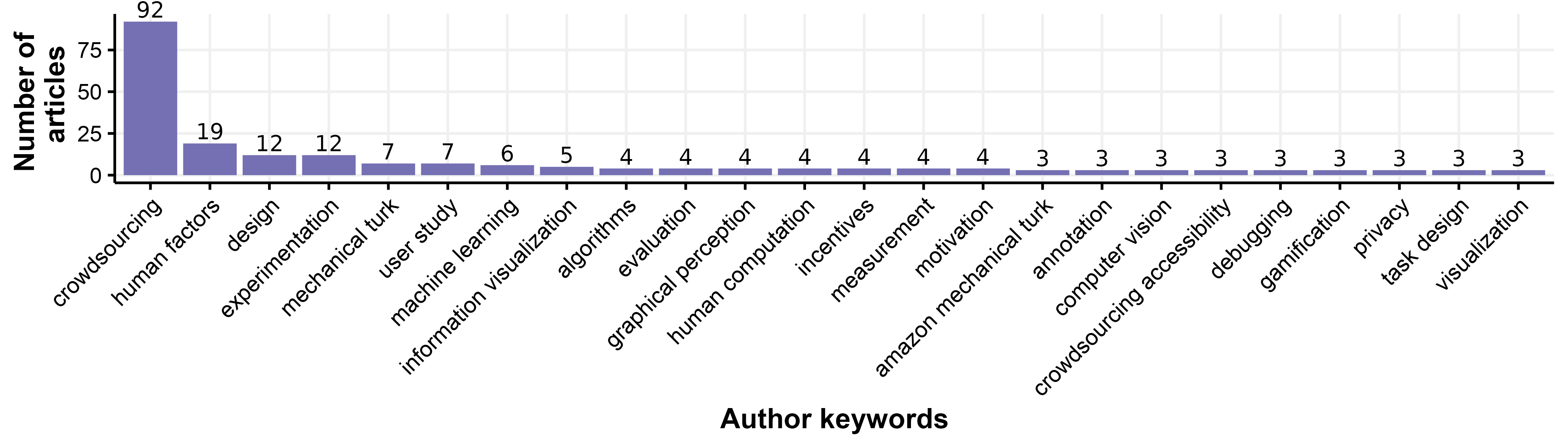}%
\caption{Bar chart of author-provided keywords appearing in the analyzed papers, including only keywords mentioned at least three times.}
    % \Description{Word cloud created from author-provided keywords in articles reporting crowd pilot studies.}
\label{fig:keywords}%
\end{figure}

Analyzing the Computing Classification System (CCS) concepts provided by the authors,
    % (see \autoref{tab:ccs-concepts}),
we find the bulk of the articles were classified as research on  Human-Centered Computing ($n=47$),
HCI ($n=45$),
Information Systems ($n=26$),
Computing Methodologies ($n=13$), and
Applied Computing ($n=7$).
A similar human-centered pattern is notable in the author-provided keywords (see \autoref{fig:keywords}) which revolve around human factors, design, user studies, and experimentation involving human subjects.
% -----
% -----

The corpus includes 770 affiliations (including double-affiliations) of authors from academia and industry.
    % (see \autoref{fig:authors}).
Authors are predominantly affiliated with universities ($n=607$),
with Carnegie Melon University (42 authors),
Stanford University (26 authors),
the University of Washington (24 authors),
and the University of Waterloo (24 authors)
having the most author affiliations in our corpus of literature.
A smaller number of authors are from industry ($n=89$) and other institutions (e.g., colleges and research institutes; 74 authors).
Among the institutions from industry are Microsoft (32 authors), Google (10 authors), and Disney Research (7 authors).

The corpus provides us a source of documentation on the state of practice of crowd pilot study reporting in the HCI and crowdsourcing literature which we analyzed to answer research questions RQ1~--~RQ4.

% --------------------
\subsection{RQ1: Why are Crowd Pilot Studies Typically Conducted?}%
\label{sec:RQ1}%
% --------------------
%
Most crowd pilot studies in the literature are formative studies conducted during the design or development phase ($n=143$, \pt{143}\%).
About 15\% of the crowd pilot studies are summative studies ($n=26$, \pt{26}\%) conducted to evaluate or validate a prototype, proof of concept, or an idea.
Three quarters of the articles in our corpus report crowd pilot studies only in passing --- in a few sentences, footnotes, or short paragraphs (128~articles, \pt{128}\%).

We classified the articles based on the amount of space a crowd pilot study is given in the article and the type of crowdsourcing study (formative versus summative).
    In this classification, `in passing' refers to articles which mention the pilot study only in few sentences, footnotes, or short paragraphs. `Detailed reporting' denotes articles which dedicate a larger amount of space (e.g., a full section) to the pilot study. Finally, `main study' refers to articles in which the entire article is considered as being a pilot study.
Based on this classification,
we find there are five different types of crowd pilot studies with varying levels of prevalence in the literature (as depicted in  \autoref{fig:pilotreporting}):%
% -----
\begin{figure}[!h]%
\centering%
\includegraphics[width=.5\linewidth]{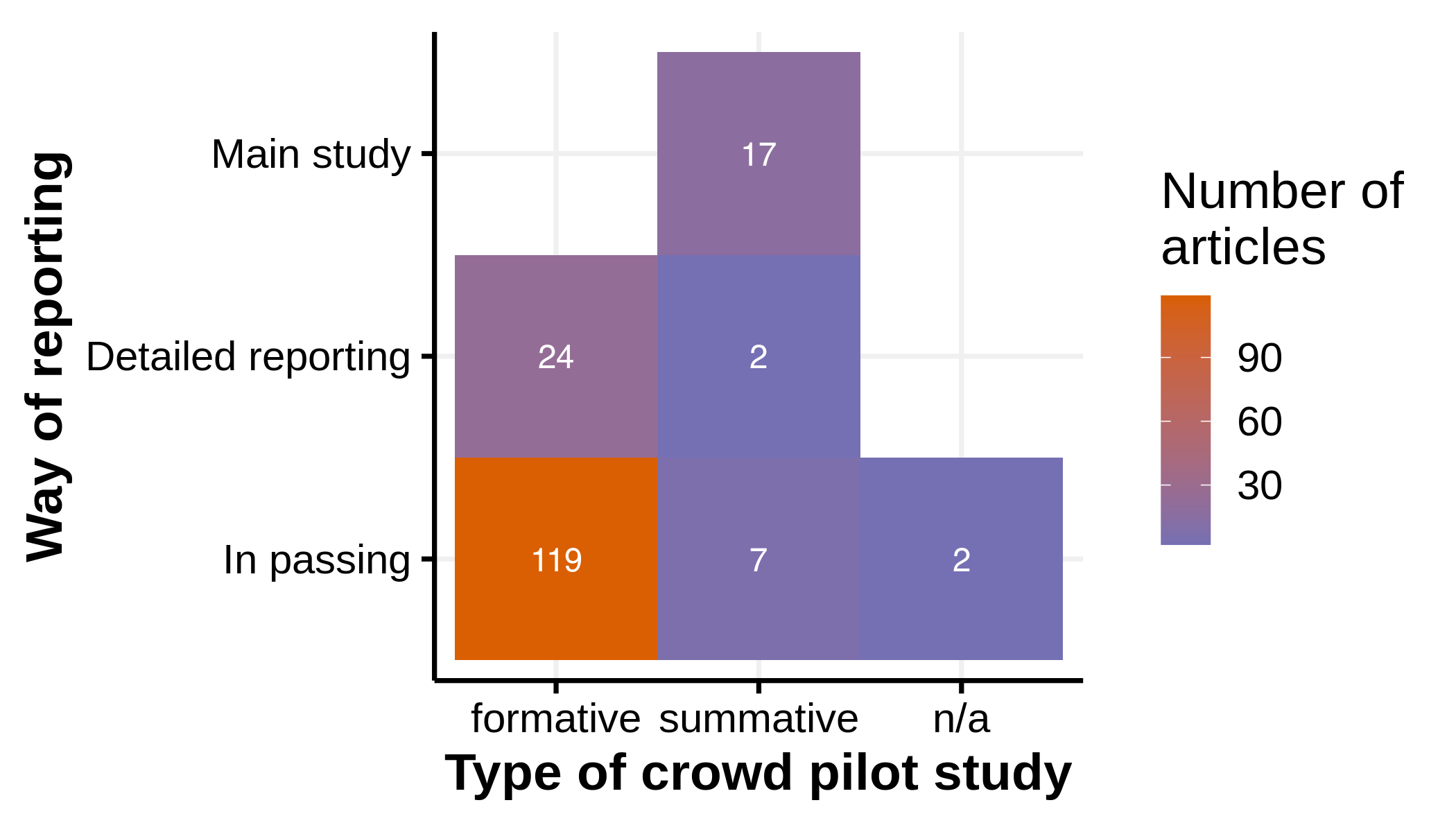}%
\caption{Types of crowd pilot studies and ways of reporting crowd pilot studies in the literature.}%
    % \Description{Types of crowd pilot studies and ways of reporting crowd pilot studies in the literature.}%
\label{fig:pilotreporting}%
\end{figure}%
% -----

\begin{itemize}%
    \item
    \textit{Formative crowd pilot study, mentioned in passing}
    ($n=119$, \pt{119}\%):
    Over two thirds of the articles in our corpus contain formative crowd pilot studies that are mentioned in passing. These articles are by far the largest group in the literature.
    The articles in this group are often opaque in how the crowd pilot study is being reported and details about the crowd pilot study are often not provided. As this is the largest group of articles, we analyze these articles in more detail in % the remainder of this section and in 
    Section~\ref{sec:wording}.% sec:RQ2

    \item
    \textit{Formative crowd pilot study, detailed reporting}
    ($n=24$, \pt{24}\%): 
    This type of article  devoted more than just a few sentences to the crowd pilot study. 
        For instance, \citet{2810188.2810190.pdf} report on a series of three formative pilot studies in which the authors iterated on parameters related to the design of the task and campaign (such as the task design, task instructions, task reliability, task accuracy, task difficulty, and worker behavior) to inform a crowdsourcing campaign. % \citep{2810188.2810190.pdf}.    

    \item
    \textit{Summative crowd pilot study, pilot is main study}
    ($n=17$, \pt{17}\%):
    Several articles presented a summative crowd pilot study as main study of the article. This type of study is being conducted to test the feasibility, provide a proof of concept, or evaluate and validate a system.
    A representative article is the work by \citet{2470654.2470686.pdf} who established the feasibility of using crowds for design feedback in the classroom.
    Other examples are 
    the technical evaluation of the VidQuiz system by \citet{3274311.pdf} and the work by 
    \citet{2772879.2772947.pdf} who mention pilot studies with the task allocation system of a disaster response system.
    %~\citep{2772879.2772947.pdf}.

    \item
    \textit{Summative crowd pilot study, detailed reporting}
    ($n=2$, \pt{2}\%):
    Only very few articles contained summative pilot studies reported in detail (i.e., in a separate section of the article).
    These studies were being conducted to validate the design and functionality of systems or to show the generalizability of the system.
        \citet{3340531.3411863.pdf} conducted a pilot study with the prototype of a spatial crowdsourcing system. This pilot study was reported after the main results section (titled \textit{``performance evaluation''}). % \citep{3340531.3411863.pdf}.
        \citet{2348283.2348400.pdf} used a crowd pilot study to demonstrate that their game generalizes to other domains. % \citep{2348283.2348400.pdf}.

    \item
    \textit{Summative crowd pilot study, mentioned in passing}
    ($n=7$, \pt{7}\%):
    A few articles mentioned a summative pilot study in passing.
    % Articles in this group are less coherent in why the crowd pilot study was being conducted.
    Some articles in this group report the crowd pilot study in the context of evaluating a system, demonstrating a proof of concept, or validating the feasibility.
        \citet{3126594.3126648.pdf}, for instance, report that a
        small team of participants conducted a pilot experiment. % \citep{3126594.3126648.pdf}.
        \citet{CSI-SE.2017.2.pdf} mention that a \textit{``set of pilot runs''} was executed \textit{``to ensure the feasibility of the study design''} (p. 34). % \citep{CSI-SE.2017.2.pdf}.
        \citet{3394978.pdf} pilot tested their CrowdUI system to ensure the system's functionality before the main study. % \citep{3394978.pdf}.
        Similarly, \citet{2541288.pdf} launched a pilot study with the intention of verifying the quality of results before deploying the main study. % \citep{2541288.pdf}.
    Other articles validate that a task produces the intended results, such as \citet{2464464.2464482.pdf} who compared the results of a pilot study with the main results of their article, finding no differences. % \citep{2464464.2464482.pdf}.
    % In very few instances, articles reported in passing on the pilot study in a summative sense to support the evaluation. For instance, \citeauthor{3411764.3445557.pdf} mention that a pilot version of their study was conducted and the authors \textit{``believe [their] results are robust to pseudoreplication''}~\citep{3411764.3445557.pdf}.
    % \citep{3359238.pdf} mention that their \textit{``pilot and actual studies demonstrated improved crowd analysis quality when setting an explicit threshold in a numerical rating task.''}

    \item
    \textit{Other articles:} Two articles (\pt{2}\%) could not be assigned to being formative or summative due to insufficient information in the article.
    These two instances are an article in which a pilot study is mentioned in the acknowledgments section~\citep{3411764.3445507.pdf}
    and the article by ~\citet{3130931.pdf} in which the authors state that crowd workers were excluded if they had \textit{``previously participated in any of [the authors'] pilot studies''} (p. 10).
\end{itemize}

Digging deeper into the details reported about the crowd pilot studies,
we find that task design and crowdsourcing campaign are the most commonly reported reasons for undertaking a crowd pilot study in the literature.
    % (see \autoref{fig:Why}).
We split our further analysis into three parts related to the motivation for running the pilot study: the design of the task, the crowdsourcing campaign, and other reasons.

% -----
% \begin{figure}[!h]%
% \centering%
% \includegraphics[width=.7\linewidth]{figures/images/pWhy.png}%
% \caption{Reasons for conducting the crowd pilot study.}%
%     \Description{Reasons for conducting the crowd pilot study.}%
% \label{fig:Why}%
% \end{figure}%
% -----

\subsubsection{Task design related reasons for conducting crowd pilot studies ($n=101$, 59.1\%)}% \pt{101}\%
The design of crowdsourced tasks is a central component of crowdsourcing and includes factors affecting the performance of tasks and quality of results, such as, for instance, the task design, the usability of the user interface, and the clarity of the task instructions.
Authors in our corpus often mentioned iterative experimentation aiming to improve the task design, but without providing details.
    \citeauthor{2786451.2786492.pdf}, for instance, mention that \textit{``[p]ilots were run for optimizing the microtasks settings in terms of cost, amount of judgments and task design''} (p. 2). % \citep{2786451.2786492.pdf}.
    \citet{3173574.3173806.pdf} report that they \textit{``iterated extensively in pilot studies with crowd workers to strike a balance between simplicity (avoid complex or numerous instructions) and effectiveness (make the layout better)''} (p. 4). % \citep{3173574.3173806.pdf}.

% -----
\begin{figure}[!h]%
\centering%
\includegraphics[width=.65\linewidth]{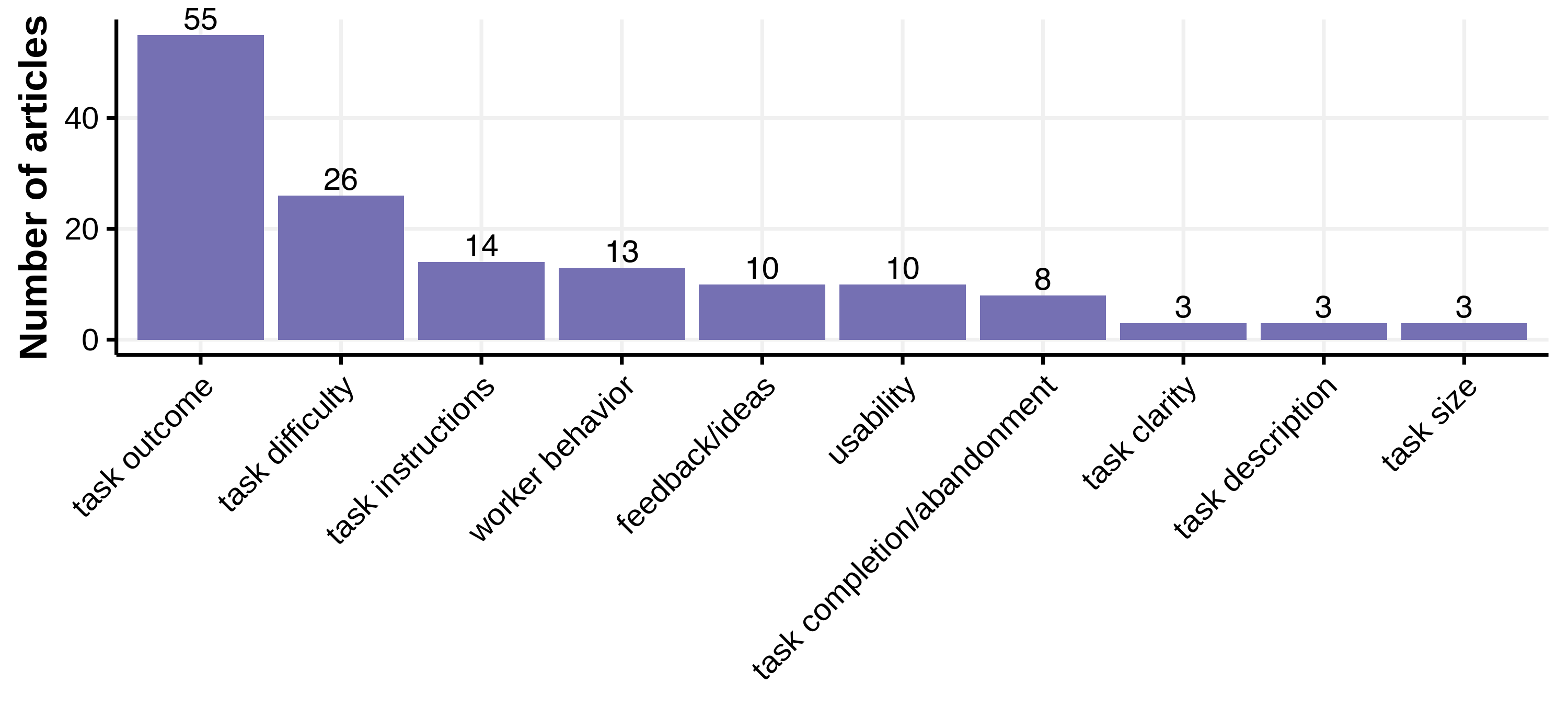}%
\caption{Task design related reasons for conducting crowd pilot studies reported in the literature.}%
    % \Description{Task design related reasons for conducting crowd pilot studies reported in the literature.}%
\label{fig:WhyDetailsTaskDesign}%
\end{figure}%
% -----

If details about the task design are mentioned in the article, we find it is most common for authors to report on the outcome of the crowd pilot study (see \autoref{fig:WhyDetailsTaskDesign}). Under this category, we subsume any information reported by authors about the crowd pilot study's results, performance, accuracy, validity, and quality.
    \citet{3366423.3380063.pdf}, for instance, determined a similarity threshold \textit{``[t]hrough a pilot study''} from the accuracy of results. % \citep{3366423.3380063.pdf}.
    \citet{2556288.2556986.pdf} \textit{``learned from pilot runs that longer video segments lead to lower annotation accuracy [...] and slower responses on Mechanical Turk''} (p. 4021). %\citep{2556288.2556986.pdf}.
    The task outcome was the most often reported factor related to task design ($n=55$, \pt{55}\%), used both in formative and summative crowd pilot studies.

Crowd pilot studies were also often conducted to assess the difficulty of the task during the formative design stage ($n=26$, \pt{26}\%).
    For instance, \citet{3306618.3314270.pdf} reported that \textit{``in pilot experiments,''} workers were unfamiliar with items presented to them.
    The solution by the authors was to use custom qualifications.
        A qualification is ``a set of questions [...] that the worker must answer to qualify and therefore work on the assignments~\citep{10.1.1.149.6649.pdf}.
    Similarly, \citet{3340531.3412782.pdf} determined \textit{``[i]n pilot experiments''} that the \textit{``task does not require expert workers, so we just required workers to have at least 100 previously approved HITs''} (p. 3051). % \citep{3340531.3412782.pdf}.
    Other authors carried out a more rigorous process to avoid systematic biases from seeping into the collected data~\citep{hube2019understanding}. 
For instance,  \citet{3399715.3399827.pdf} systematically experimented with different difficulty levels in a pilot study ``\textit{to identify the correct level of difficulty to avoid floor and ceiling effects}'' (p. 5). % \citep{3399715.3399827.pdf}.

Another reason for conducting the crowd pilot study is the iterative design of task instructions ($n=14$, \pt{14}\%), including to improve a task's intelligibility (e.g., \citet{3375187.pdf}) and clarity (e.g., \citet{5271-Article_Text-8372-1-10-20190926.pdf}).
However, many authors were not specific on how the task instructions were iterated on.
    For instance, \citet{2555691.pdf} simply reported that the ``\textit{exact instructions we gave workers evolved over the course of several pilot experiments}'' (p. 18). % \citep{2555691.pdf}.

Analyzing worker behavior and preferences was another~-- although with 13 articles (\pt{13}\%) less common~-- reason for conducting crowd pilot studies.
    \citet{3366424.3383546.pdf}, for instance, measured parameters of a worker model \textit{``according to a pilot task on Figure Eight''} (p. 225). % \citep{3366424.3383546.pdf}.
    \citet{3290605.3300750.pdf} reported on pilot experiments that \textit{``revealed people had a strong preference to use manual control''} (p. 4). % \citep{3290605.3300750.pdf}.
    \citet{3328906.pdf} investigated the workers' behavior during the crowd pilot study, including the response rate, and reported that workers already adopted the tasks as a habit during the crowd pilot study. % \citep{3328906.pdf}.

Some authors elicited open-ended feedback in the crowd pilot study and ideas for improving the study during the formative design phase ($n=10$, \pt{10}\%).
    For instance, \citet{2897370.pdf} wrote that \textit{``[t]o better inform the interface, we conducted a pilot study with 5~non-expert workers and asked them to rate the appearance of the marks after they finished their tasks''} (p. 9). % \citep{2897370.pdf}.
    \citet{2984511.2984578.pdf} reported that according to feedback from the pilot studies, their \textit{``approach was intuitive and matched users’ expectations well''} (p. 613). % \citep{2984511.2984578.pdf}.
    \citet{2818048.2819979.pdf} used the exit survey on CrowdFlower during the crowd pilot study to monitor the workers' satisfaction with the payment to \textit{``ensure fair worker treatment''} (p. 265). %\citep{2818048.2819979.pdf}.
    % Besides the use of qualifications,
    However, CrowdFlower's exit survey was rarely used to specifically support the aims of crowd pilot studies. 

Usability ($n=10$, \pt{10}\%) and the task abandonment~\citep{han2019all,han2019impact} or completion rate ($n=8$, \pt{8}\%) were also mentioned in articles.
    \citet{2872427.2883035.pdf}, for instance, reported that the \textit{``iterative design resulted in substantial usability improvements''} (p. 135) % \citep{2872427.2883035.pdf}
    and \citet{2998181.2998311.pdf} observed the task completion rate of 900~workers in \textit{``a pilot study [with] over 900~workers in Sept 2015. From that study, [the authors] observed that 15\% of the accepted tasks are not completed by the crowd workers''} (p. 907). % \citep{2998181.2998311.pdf}.
The task description ($n=3$, \pt{3}\%) and the task clarity ($n=3$, \pt{3}\%) were only explicitly mentioned in a few articles, although these two items are implicitly part of the design of the task and its instructions.
Another equally less frequently reported reason for conducting the crowd pilot study includes empirically determining the optimal size of a task ($n=3$, \pt{3}\%).
    For instance, \citet{2598510.2598514.pdf} \textit{``tested a variety of gameplay settings''} to determine the optimal number of items included in a task to \textit{``not cause fatigue''} (p. 708). % ~\citep{2598510.2598514.pdf}.

\subsubsection{Campaign design related reasons for conducting crowd pilot studies ($n=58$, 33.9\%)}% \pt{58}\%
The design of a crowdsourcing campaign includes parameters necessary for launching the campaign, such as the number of tasks assigned to workers or batch sizes, the average task completion time, and the task pricing. These factors have been shown to influence task outcomes~\citep{difallah2014scaling,difallah2015dynamics,cheng2015measuring}. 
    % (see \autoref{fig:WhyDetailsCampaign}).

% -----
% \begin{figure}[!h]%
% \centering%
% \includegraphics[width=.6\linewidth]{figures/images/pWhyDetailsTaskParameters.png}%
% \caption{Campaign related reasons for conducting crowd pilot studies reported in the literature.}%
%     \Description{Campaign related reasons for conducting crowd pilot studies reported in the literature.}%
% \label{fig:WhyDetailsCampaign}%
% \end{figure}%
% -----

Related to the campaign design, crowd pilot studies are often conducted to empirically determine the price of the crowdsourced task ($n=35$, \pt{35}\%).
    As mentioned in the introduction, the task price is typically estimated from the workers' average task completion times in a crowd pilot study. We see evidence for this practice in our literature corpus.
    % (cf. \autoref{fig:WhyDetailsCampaign}).
% Authors often mention that the average task completion time was used to estimate the price of tasks.
% In nine articles, the estimation of the task price from the average completion times was the only information related to task design provided by the authors.
Typical ways of reporting this information include, for instance,
    \citet{2513383.2513448.pdf} who reported that workers were \textit{``paid \$0.75 per HIT (\$0.047--0.054 per labeling task); which was decided based on the task completion time in pilot studies (e.g., approximately \$0.10 per minute)''} (p. 6). %~\citep{2513383.2513448.pdf}.
    Another way was to provide information on a target hourly rate, such as \citet{3173574.3174216.pdf} who mentioned that
    \textit{``[b]ased on piloting, we paid participants \$2 for participation, for a target rate of \$8/hour''} (p. 5). % ~\citep{3173574.3174216.pdf}.
    Similarly, \citet{3503516.3503533.pdf} mention in a footnote that \textit{``[b]ased on [a] pilot experiment''} the hourly pay was \textit{``equivalent to US\$13.5 per hour on average''} (p. 3). %~\citep{3503516.3503533.pdf}.
    A slightly more extensive report % using the technique of estimating the task price from average task completion times
    was given by \citet{3397271.3401112.pdf} who \textit{``performed several small pilots of the task, and after measuring the time and effort taken to successfully complete it, we set the HIT reward to \$1.5. This was computed based on the expected time to complete it and targeting to pay at least the US federal minimum wage of \$7.25 per hour''} (p. 441). %~\citep{3397271.3401112.pdf}.
% The average time spent was calculated and the pay for the main experiment was set to \textit{``a maximum of \$4.00 to meet the federal minimum wage''}.
% Bonuses were also tested in crowd pilot studies by the authors.

Besides the very common combination of task price estimated from average task completion times, some authors also empirically determined other campaign related parameters in crowd pilot studies.
This includes determining a time limit for the task \citep{3173574.3173884.pdf,2505515.2505695.pdf,2598510.2598514.pdf,13273-Article_Text-16790-1-2-20201228.pdf,13296-Article_Text-16813-1-2-20201228.pdf,3242587.3242598.pdf,2600428.2609577.pdf,3411764.3445399.pdf}
and determining the optimal sample size
\citep{5268-Article_Text-8369-1-10-20190926.pdf,3269206.3271779.pdf,7470-Article_Text-10869-1-10-20200930.pdf,2766929.pdf,5264-Article_Text-8365-1-10-20190926.pdf,3290605.3300742.pdf,3411764.3445344.pdf,2740908.2745401.pdf} for the main study.
Only few crowd pilot studies  involved qualifications for a task.
    \citet{3306618.3314270.pdf} used a \textit{``qualification exam''} to identify well-performing workers based worker accuracy. % \citep{3306618.3314270.pdf}.
    \citet{3340531.3412782.pdf} used results from a crowd pilot study to determine the minimum number of previously approved HITs for workers. % \citep{3340531.3412782.pdf}.
    \citet{5268-Article_Text-8369-1-10-20190926.pdf} analyzed the geographic location of workers in the crowd pilot study to identify countries for the main study. % \citep{5268-Article_Text-8369-1-10-20190926.pdf}.
    \citet{2993148.2993200.pdf} used a quiz on CrowdFlower to filter workers. % \citep{2993148.2993200.pdf}.
    \citet{3321700.pdf} ruled out the use of qualifying questions through crowd pilot studies to avoid an increase in attrition rate. % \citep{3321700.pdf}.

A common way of controlling the quality in crowdsourcing studies are gold questions (i.e., questions for which the answer is known) \citep{DanielCSUR2017.pdf}.
One way of developing and verifying gold standard questions would be through crowd pilot studies. However, only few articles mentioned gold standard questions in the context of crowd pilot studies.
    \citet{13287-Article_Text-16804-1-2-20201228.pdf}
    found an \textit{``inexplicable problem''} with the gold judgments and subsequently abandoned the use of the gold dataset in favor of another dataset. % \citep{13287-Article_Text-16804-1-2-20201228.pdf}.
    \citet{2740908.2745401.pdf} used a crowd pilot study with seven MTurk workers to verify that the quality of work done is comparable to trained experts, concluding that \textit{``judgements from
    20 workers on Mturk can serve as the gold standard data set''} (p. 403).
    % \citep{2740908.2745401.pdf}.
    \citet{2661829.2661918.pdf} mention that \textit{``[s]mall pilot experiments were carried out while developing the design of the HIT''} which included \textit{``gold labels''} (p. 323). % \citep{2661829.2661918.pdf}.
    This gold standard was taken from existing corpora and not verified.
    Last, \citet{2810188.2810190.pdf} went one step further and presented experimentation on gold standards in one of their crowd pilot studies. The authors found % , for instance,
    that \textit{``the gold standard proved to be too restrictive''} and \textit{``gold tests and majority voting produced approximately the same acceptance results''} (p. 29). %     \citep{2810188.2810190.pdf}.

\subsubsection{Other reasons for conducting a crowd pilot study ($n=72$, 42.1\%)}% \pt{72}\%
In the above, the majority of crowd pilot studies are conducted for formative reasons with the aim of iteratively designing a crowdsourced task in rapid fashion with a small set of participants.
We found that another formative reason for conducting a crowd pilot study is collecting data for the main study (20 articles, \pt{20}\%).
This category of articles can be split in articles which conducted the crowd pilot study with the sole purpose of collecting data (9 articles) and articles that conducted the crowd pilot study also for other purposes (11 articles).
    \citet{13072-Article_Text-16589-1-2-20201228.pdf}, for instance, collected \textit{``pre-generated solutions represented common wrong solutions that were submitted by participants (as determined in a pilot study)''} (p. 5). % \citep{13072-Article_Text-16589-1-2-20201228.pdf}.
    \citet{2818048.2835201.pdf} conducted a pilot study to collect constraints for a design task. % \citep{2818048.2835201.pdf}.
In these articles, the collected data was then used in the main experiment or study of the article.
    For instance, \citet{2998476.2998498.pdf} crowdsourced a dataset of tagged tweets in a crowd pilot study which was then used as input for a machine based classifier, \textit{``thereby making the classifier emotionally intelligent''} (p. 3). %~\citep{2998476.2998498.pdf}.

Another reason for conducting the crowd pilot study was the iterative design of a study or an experiment.
% study design/experiment design
This purpose of a crowd pilot study was explicit in 11~articles (\pt{11}\%), although we believe this motive is implicitly present in many articles, such as the ones reporting on iteratively designing a task in the context of an experiment or the design of a system.
As a single outstanding instance, \citet{3359227.pdf} used a crowd pilot study to qualify and recruit participants for the main study. %~\citep{3359227.pdf}.

As observed through our findings, crowd pilot studies are carried out for a broad range of compelling reasons -- reasons that others who partake in carrying out crowdsourcing studies are very likely to face. It is therefore, important to understand how and at what level of detail crowd pilot studies are reported in literature.

% --------------------
% \subsection{In what sections of their article are authors reporting pilot studies?}%
\subsection{RQ2: How are Crowd Pilot Studies Typically Reported?}%
\label{sec:RQ2}%
% --------------------
%
In the previous section, we already provided some examples on how authors reported results of crowd pilot studies  in the scholarly literature.
In this section, we go in-depth and investigate how authors report crowd pilot studies. We analyze which terms and phrasing authors use to refer to pilot studies and how consistent they are in their wording (Section \ref{sec:wording}) as well as in which section of the article crowd pilot studies are being reported (Section \ref{sub:section}).

% --------------------
% \subsubsection{How are crowd pilot studies being referred to in the literature?}%
\subsubsection{How do authors refer to pilot studies?}%
\begin{figure}[!h]%
\centering%
\includegraphics[width=.75\linewidth]{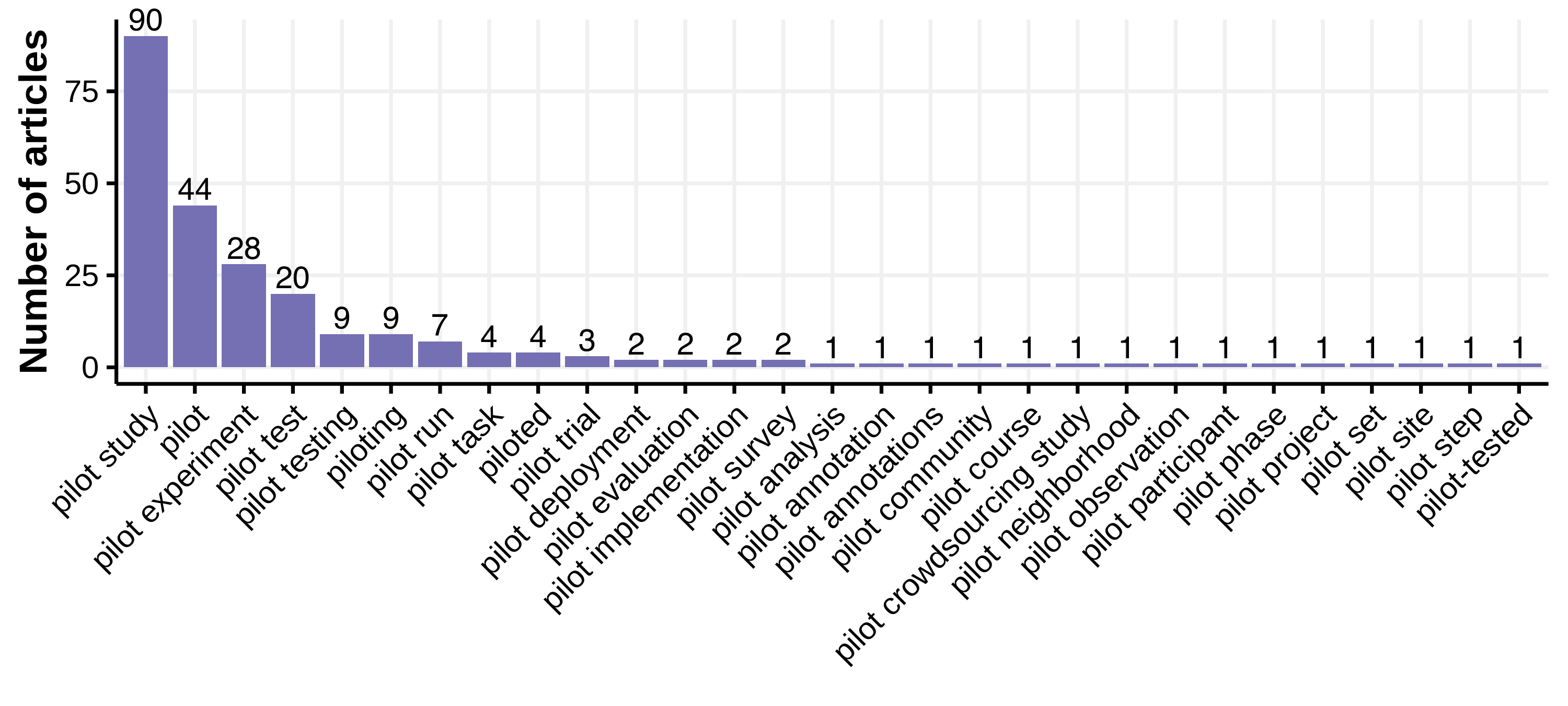}%
\caption{Different terms used to refer to pilot studies in the articles.
Note that some authors used multiple terms to refer to crowd pilot studies within their article.}%
    % \Description{Different terms used to refer to pilot studies in the articles. Some authors used multiple terms to refer to crowd pilot studies within their article.}%
\label{fig:wording}%
\end{figure}

\label{sec:wording}%
% --------------------
%
Different terms are being used in the scholarly literature to denote the provisional nature of pilot studies (see \autoref{fig:wording}).
By far the most common term used by authors was `pilot study' (90 articles, \pt{90}\%), followed by `pilot' ($n=44$, \pt{44}\%), `pilot experiment' ($n=28$, \pt{28}\%) and `pilot test' ($n=20$, \pt{20}\%).
Most of the terms chosen by authors are nouns, although few authors also use `pilot' as a verb (e.g., \textit{``piloted''}), present participle (\textit{``piloting''}), and gerund (\textit{``pilot testing''}).
% -----
% -----

Given that the vast majority of articles reported on pilot studies only in passing, a surprisingly low number of articles ($n=5$, \pt{5}\%) referred to the crowd pilot study as an \textit{``informal''} study \citep{2528394.2528396.pdf,3178876.3186031.pdf,2858036.2858279.pdf,2642918.2647403.pdf,2903138.pdf}.
This informal study provided authors an \textit{``informal sense''} of worker behavior in the context of an \textit{``open-ended exploration''} \citep{2528394.2528396.pdf} as well as support in the design of tasks to \textit{``understand how different interfaces affected crowd performance''} \citep{2642918.2647403.pdf}.

% --------------------
% \subsection{How consistent are authors in their wording within the article?}%
% \label{sub:wordingconsistency}%
% --------------------

We find that about two thirds of the articles ($n=115$, \pt{115}\%) are internally consistent in how they refer to the pilot study within the article. In these articles, authors used only one single term to refer to the pilot study.
% 46 articles are inconsistent in their use of wording.
In the other third of the articles, some authors used up to four different terms to refer to the pilot study (43 articles used two different terms, 12 articles used three terms, and one article used four terms).
The most common combinations among the articles using multiple terms are
    `pilot study' and `pilot'	(n=12),
    `pilot test' and `pilot study'	(n=78),
    `pilot study' and `pilot experiment'	(n=3),
    `pilot experiment' and `pilot'	(n=3), and
    `pilot study' and `pilot run'	(n=3).
% Among the articles which used two terms, pilot study (n=24), pilot (n=17), and pilot test (n=8) were most common.
% Among the articles that used three terms, pilot study was also most common (n=8) followed by pilot (n=6) and pilot experiment (n=5).

\begin{figure}[!h]%
\centering%
\includegraphics[width=.75\linewidth]{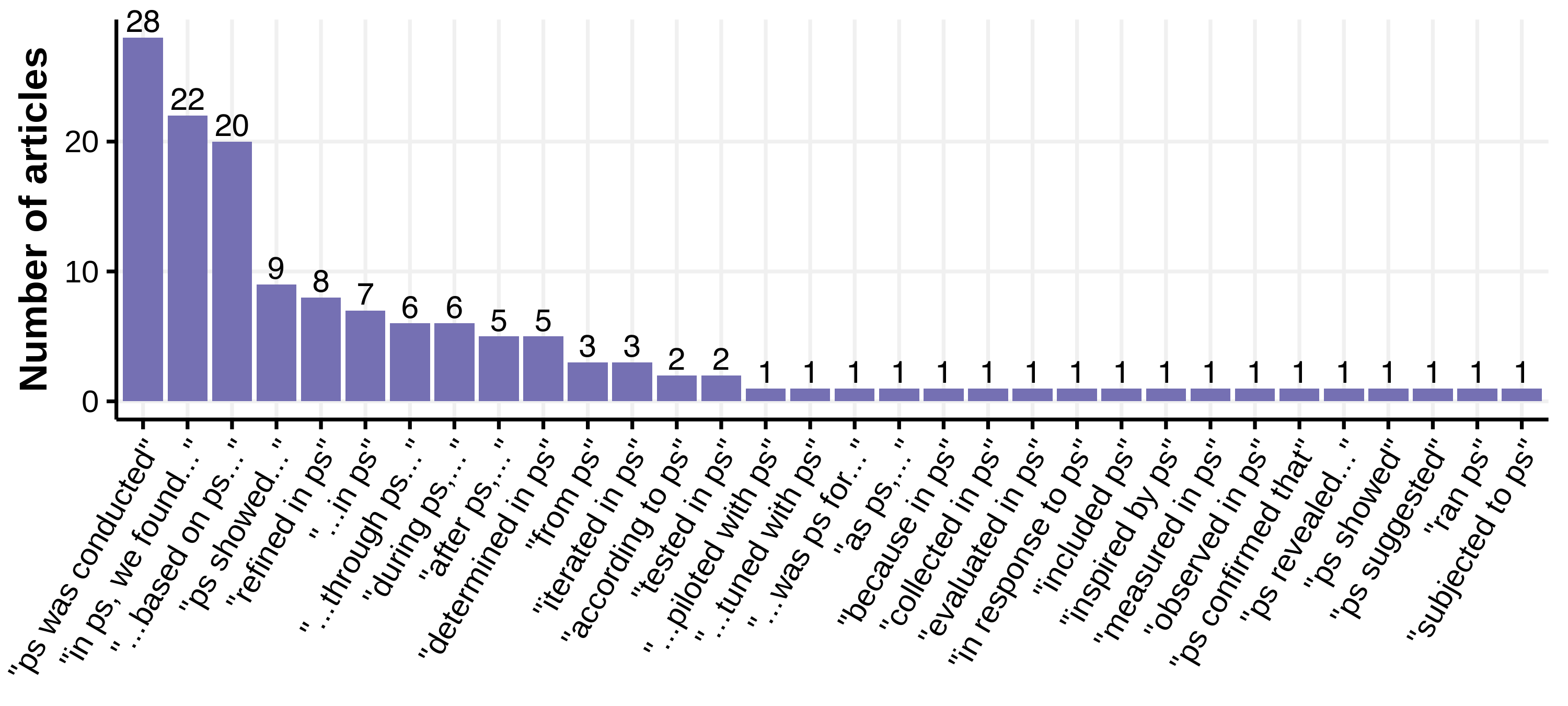}%
\caption{Phrasing used to report on the results of pilot studies (abbreviated `ps' in this figure) among articles which mention formative crowd pilot studies.}%
    % \Description{Phrasing used to report on the results of a pilot study (abbreviated `ps' in this figure) among articles which mention formative crowd pilot studies.}%
\label{fig:phrasing}%
\end{figure}

% --------------------
% \subsubsection{How do authors phrase their reports of crowd pilot studies?}%
\label{sub:phrasing}%
% --------------------
In formative crowd pilot studies mentioned in passing,
the term `pilot study' is often used as a blanket statement to justify design decisions without providing details about the crowd pilot study.
For this reason, we analyzed the phrasing used by authors of these crowd pilot studies in more detail (see \autoref{fig:phrasing}).

% -----
% -----

The most common way in which authors mention formative crowd pilot studies is by stating that a crowd pilot study was conducted, followed by selected details about the outcome of the pilot study.
    \citet{2661829.2661918.pdf}, for instance, mention that 
    \textit{``[s]mall pilot experiments were carried out while developing the design of the HIT,''} followed by details about the HIT, the assignment of the HIT to workers, and the price of the HIT. % \citep{2661829.2661918.pdf}.
The phrase `in a pilot study, we found' (or close derivations thereof) is also common.
    For instance, \citet{2513383.2513448.pdf} report that \textit{``[i]n early pilot studies, we found that users would get disoriented''} (p. 6). %~\citep{2513383.2513448.pdf}.
It is also very common for authors to derive design decisions `based on a pilot study.' This phrasing was often used to refer to the estimation of the task price from average (or in some cases median) task completion times.
    For instance,
    \citet{3479550.pdf} report that workers were offered \$0.50 per rating \textit{``based on the median time taken on a pilot task''} (p. 10). % \citep{3479550.pdf}.
    Similarly, \citet{3359238.pdf}
    estimated \textit{``the time needed for each microtask based on pilot studies''} (p. 7). %  \citep{3359238.pdf}.
Some other phrases used among authors include that crowd pilot studies `showed,' `revealed,' or `demonstrated' some specific results and
    that parameters were iteratively `refined in pilot studies.'

Next, to draw further insights about pilot studies based on the context in which they are described, we explored sections of articles in which they are reported.

% --------------------
\subsubsection{In which section are pilot studies reported?}%
\begin{figure}[!h]%
\centering%
\includegraphics[width=.7\linewidth]{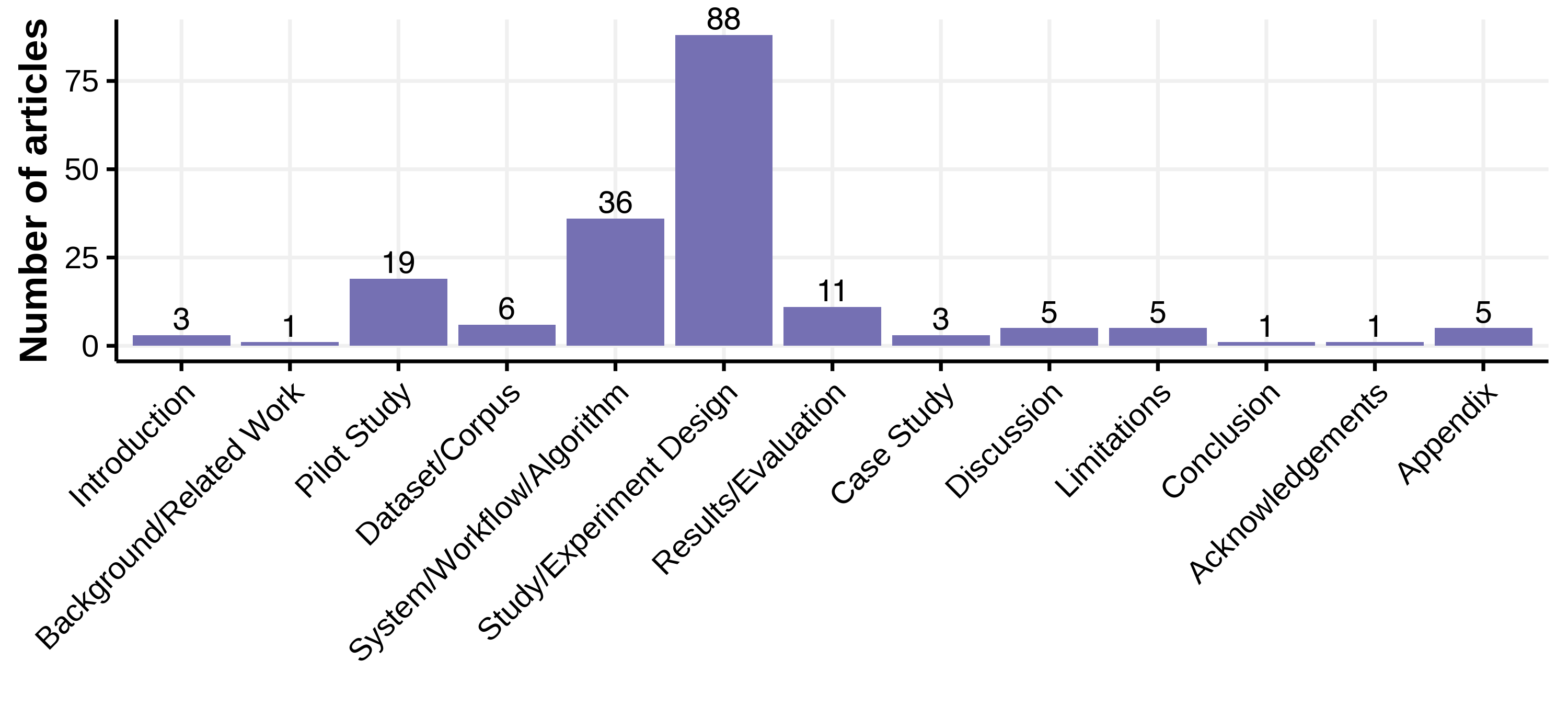}%
\caption{Sections in which authors report the results of their pilot studies.}%
    % \Description{Sections in which authors report the results of their crowd pilot studies.}%
\label{fig:section}%
\end{figure}

\label{sub:section}%
% --------------------
%
We analyzed in which section authors report on the crowd pilot study, using a closed-coding approach. Our initial coding scheme reflected the standard structure of academic articles (i.e., Introduction, Related Work, Method, Results, Discussion, Conclusion, Appendix). However, the codes were slightly modified after one iteration of coding to better accommodate differences in the methodological approaches used in the articles. The result of the coding is depicted in \autoref{fig:section}.

% -----
% -----

We find the majority of articles ($n=149$, \pt{149}\%) report on the crowd pilot studies in sections related to the methodology. These sections include
    the study design or experiment design ($n=88$, \pt{88}\%),
    the system design (or related sections; $n=36$, \pt{36}\%),
    dataset creation ($n=6$, \pt{6}\%),
    as well as separate sections dedicated to the pilot study, as found in about 10\% of the articles ($n=19$).
The choice of section depends on the methodological approach taken in the article. For instance, articles that develop a novel system often mention the results of the crowd pilot study in the section on the system's design.

    % One notable difference in the pilot study was that the impact of structural features had a signifcant negative impact on Likert scales for supervisors, while here this efect appears weaker.
    % \citeauthor{3479586.pdf} mention that the payment rate was estimated \textit{``based on the time estimations gathered from our pilot study and the minimum wage of 7.25 USD/hr in the United States''} \citep{3479586.pdf}. This is then used to mention future plans for adopting a more flexible payment scheme.
    %  we will adopt more flexible payment methods such as setting the payment rate based on the time spent by workers [25, 35], or we will simply pay more than the basic estimations to leave enough margin for ensuring a fairer payment rate to workers.
    % \citep{3479586.pdf}.
    % \citeauthor{3209624.pdf} defended the choice of approach taken in the pilot study. mentioning that a different approach was considered during pilot testing \citep{3209624.pdf}.

Besides the general trend described above, some outstanding instances of articles took a different approach to reporting on the crowd pilot study.
Of the outstanding instances of articles that report on the crowd pilot study in the limitations section, we were expecting that the authors would discuss weaknesses and limitations of the crowd pilot study. Instead, the crowd pilot study was, in some cases, used to validate the results of the main study.
    For instance, \citet{13325-Article_Text-16842-1-2-20201228.pdf} mention in the limitations section that \textit{``a set of pilot runs``} was executed \textit{``to ensure the feasibility of the study design''} in an application domain to \textit{``address external threats to validity''} (p. 171). % \citep{13325-Article_Text-16842-1-2-20201228.pdf,CSI-SE.2017.2.pdf}.
\citet{3411764.3445557.pdf} discuss differences between their main study and an (independent) pilot study, reporting that the results \textit{``were fully consistent with those from a pilot version of this study that we conducted in July 2019''} and that \textit{``results are robust to pseudo\-replication''} (p. 12). %~\citep{3411764.3445557.pdf}.

\citet{3357384.3357976.pdf} conducted a crowd pilot study in the introduction section to motivate their article. % \citep{3357384.3357976.pdf}.
    \citet{2903138.pdf} mentioned an \textit{``independent''} crowd pilot study which was used to estimate the optimal price of the task % \citep{2903138.pdf}, 
    and \citet{3411764.3445557.pdf} also mentioned an independent crowd pilot study.
    However, besides these three articles, crowd pilot studies were typically not conducted as independent studies, but as integral part of the article.
On the other hand, some authors used the extended space of the appendix to report on the crowd pilot study in detail.
    For instance, \citet{3479572.pdf} report differences between the main experiment and the crowd pilot study in a separate appendix. % ~\citep{3479572.pdf}.

In the following section, we investigate in more detail what is known about the pilot studies from the reporting in the literature.

% --------------------
\subsection{RQ3: What do Crowd Pilot Studies Report?}%
\label{sec:RQ3}%
% --------------------
This section reports findings on 
which crowdsourcing platform is being used (Section \ref{sec:platform}),
how many crowd pilot studies are being conducted in each article (Section \ref{sec:pilotstudynumber}),
and which other key details are being reported about the crowd pilot study (Section \ref{sec:details}).

% --------------------
\subsubsection{Which crowdsourcing platform is being used?}%
\label{sec:platform}%
% --------------------
%
We analyzed which crowdsourcing platform is being used in crowd pilot studies.
This information was often not explicitly stated and had to be inferred from context.

% -----
% \begin{figure}[!h]%
% \centering%
% \includegraphics[width=.6\linewidth]{figures/images/pCrowdOther.png}%
% \caption{Crowdsourcing platforms and participant pools used in crowd pilot studies.}%
%     % \Description{Crowdsourcing platforms and participant pools used in crowd pilot studies.}%
% \label{fig:CSplatform}%
% \end{figure}%
% -----

Amazon Mechanical Turk (MTurk) is by far the most common crowdsourcing platform ($n=102$, \pt{102}\%) in the literature corpus. % (see \autoref{fig:CSplatform}).
Other platforms include, for instance, 
    CrowdFlower/Figure Eight (now Appen) ($n=23$, \pt{23}\%),
    Prolific \citep{3411764.3445344.pdf,3365610.3365621.pdf,3411764.3445637.pdf},
    Microworkers \citep{2810188.2810190.pdf,2660114.2660126.pdf},
    LabInTheWild \citep{2984511.2984578.pdf},
    % Toloka \citep{3437963.3441831.pdf},
    ZBJ \citep{3375187.pdf},
    Clickworker \citep{3411764.3445557.pdf}, and
    the Yahoo! crowdsourcing platform \citep{18948-Article_Text-22714-1-2-20211004.pdf}, among others.
    In about 12\% of the articles ($n=20$, \pt{20}\%), the crowd pilot study was conducted with other participant samples, such as students \citep{3021460.3021483.pdf,7471-Article_Text-10867-1-10-20200930.pdf,2998181.2998311.pdf,3287050.pdf}, citizens \citep{2517899.2517930.pdf,3209281.3209347.pdf,13322-Article_Text-16839-1-2-20201228.pdf,2145204.2145310.pdf,2598510.2598514.pdf,2858036.2858327.pdf}, and volunteers \citep{2502081.2502221.pdf,2504776.2504796.pdf,2596695.2596713.pdf,3349611.3355545.pdf,2384916.2384941.pdf,7472-Article_Text-10866-1-10-20200930.pdf,2971485.2971492.pdf,3178876.3186031.pdf}.
    In-house or custom crowdsourcing platforms were only reported in five studies (\pt{5}\%) \citep{3238147.3240727.pdf,3340531.3411863.pdf,3173574.3173850.pdf,18947-Article_Text-22713-1-2-20211004.pdf,2897370.pdf}.
        These articles include a pilot study conducted on an \textit{``indigenous crowdsourcing platform''} \citep{3238147.3240727.pdf} and an article by authors from Google who \textit{``ran numerous pilots to tune task hyper-parameters [...] sourced from contracted operators through an in-house crowdsourcing platform''} \citep{18947-Article_Text-22713-1-2-20211004.pdf},
        a study with the prototype of a spatial crowdsourcing system \citep{3340531.3411863.pdf},
        a study with a web-based platform for collecting ratings \citep{3173574.3173850.pdf},
        and a study conducted with a crowdsourcing system designed for analyzing industrial tomographic images~\citep{2897370.pdf}.
In ten articles (\pt{10}\%), neither the type of participant sample nor crowdsourcing platform was mentioned.

% --------------------
\subsubsection{How many pilot studies are being conducted in the article?}%
\begin{figure}[!h]%
\centering%
\includegraphics[width=.6\linewidth]{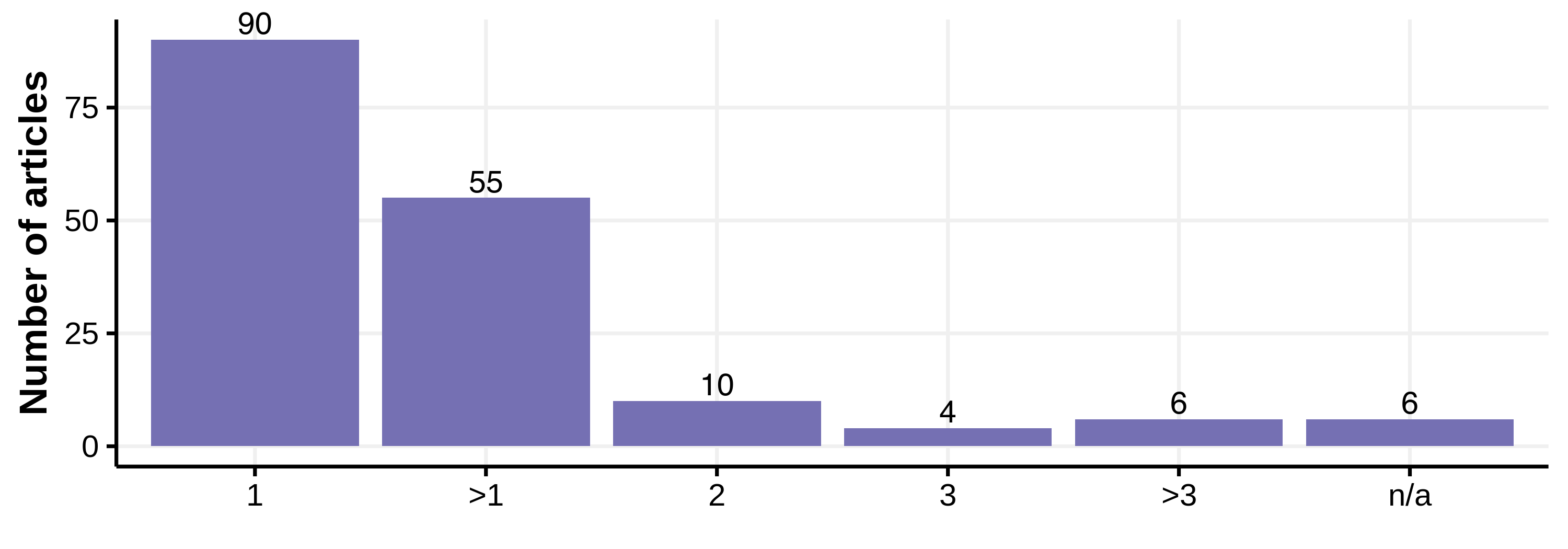}%
\caption{Number of pilot studies conducted in the article.}%
    % \Description{Number of pilot studies conducted in the article.}%
\label{fig:numpilotstudies}%
\end{figure}

\label{sec:pilotstudynumber}%
% --------------------
%
As in the previous section, the information on how many pilot studies were conducted in the article had to, in many instances, be inferred from the wording used by the authors.
In many cases, this wording was opaque and the exact number of crowd pilot studies could not be determined (see \autoref{fig:numpilotstudies}).
    For instance, some authors mentioned conducting
        `pilot studies' (e.g., \citet{3025453.3025870.pdf,2513383.2513448.pdf}),
        `pilots' (e.g., \citet{3321700.pdf,13239-Article_Text-16756-1-2-20201228.pdf}), or
        `pilot experiments' (e.g., \citet{2786451.2786492.pdf,3306618.3314270.pdf}).
From these terms, we can only infer that there was more than one crowd pilot study conducted.
Due to the use of non-descriptive terms such as
    `pilot testing' \citep{3209624.pdf,2145204.2145354.pdf,3411764.3445493.pdf}
    and `piloting' \citep{2598510.2598587.pdf,2669485.2669511.pdf,3173574.3174216.pdf}, the number of crowd pilot studies could not be determined in six articles.%
%
% -----
% -----

About half of the articles ($n=90$, \pt{90}\%) report conducting one pilot study (see \autoref{fig:numpilotstudies}).
A third of the articles ($n=55$, \pt{55}\%) report conducting more than one pilot study.
A high number of pilot studies within an article was rare ($n=6$, \pt{6}\%).
%%% give examples of those with high numbers
    \citet{5271-Article_Text-8372-1-10-20190926.pdf}, for instance, conducted \textit{``six small pilot tests''} to \textit{``ensure that the questions were clearly phrased, and of appropriate difﬁculty''} (p. 174). % \citep{5271-Article_Text-8372-1-10-20190926.pdf} and
    The highest number of pilot studies was reported by \citet{3269206.3271779.pdf} who conducted extensive experimentation in eight crowd pilot studies. % \citep{3269206.3271779.pdf}.

% --------------------
\subsubsection{Which key attributes are typically reported about crowd pilot studies?}%
\label{sec:details}%
% --------------------
%
We analyzed what authors choose to report about crowd pilot studies.
We specifically looked at three key attributes of crowd pilot studies: the number of workers participating in the crowd pilot study, the number of tasks (or any other information given in the article that would allow to determine the number of assignments to workers), and the rewards to workers.

% -----
\begin{figure}[!htb]%
\centering%
\includegraphics[width=.65\linewidth]{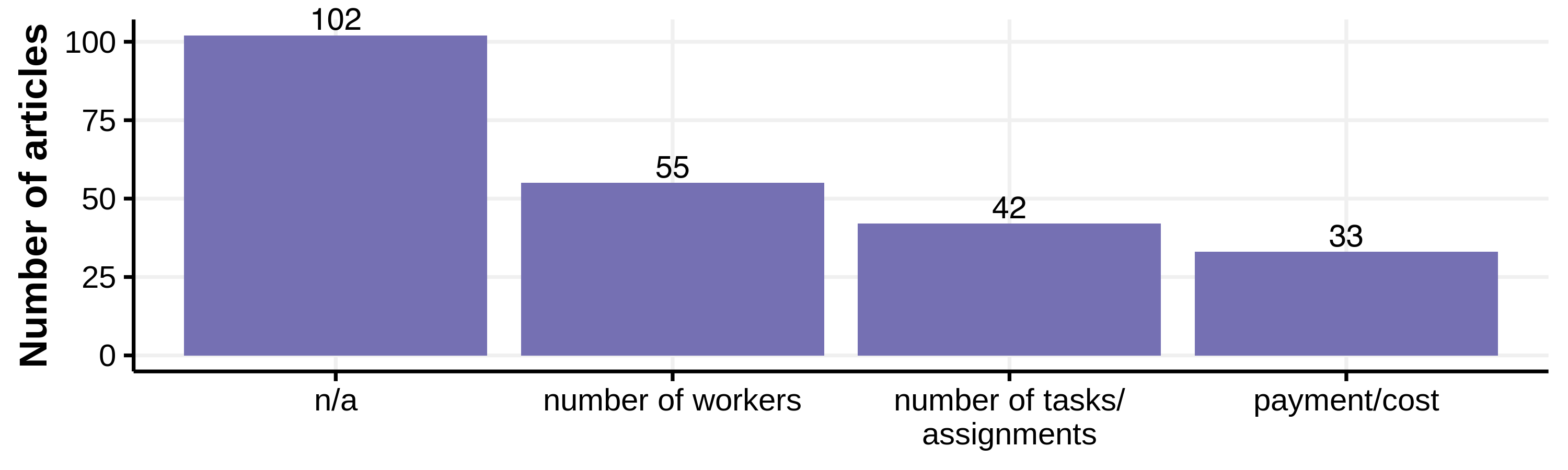}%
\caption{Key statistics reported about crowd pilot studies.}%
    % \Description{Key statistics reported about crowd pilot studies.}%
\label{fig:keystats}%
\end{figure}%
% -----

About 60\% of the articles did not provide any information on the three key statistics ($n=102$, \pt{102}\%; see \autoref{fig:keystats}).
We find authors who report one key statistic also often report other key statistics about the crowd pilot study.
Twenty-four articles (\pt{24}\%) report on all three key statistics.
    % \citeauthor{draws2021checklist}, for instance, reported 109~workers making 1994 annotations with a pay of US\$2 per task and a bonus of US\$0.50 bonus \citep{2348283.2348400.pdf}.
These articles often dedicated a full section or the entire article to the crowd pilot study.
About 40\% of the articles ($n=68$, \pt{68}\%) report at least one of the three key statistics.

About a third of the articles mentioned the number of workers participating in the crowd pilot study (55~articles, \pt{55}\%).
In these articles, the number of workers ranged from three (e.g., \citet{3365610.3365621.pdf}) to over 2,000 \citep{2502081.2502221.pdf}.
Some authors were imprecise about the number of participating workers, such as \citet{3209281.3209347.pdf} who reported \textit{``almost 40''} participants (p. 5). % \citep{3209281.3209347.pdf}.
Among the 50~articles in which we could identify or calculate the exact number of participants in the crowd pilot study, the average number of participants was 111.7 ($SD=169.9$).

The number of tasks or assignments was mentioned in 42~articles (\pt{42}\%).
This information was more difficult to analyze because some authors mentioned the number of assignments, others the number of tasks. Further complicating the analysis was the fundamental difference between the studies~-- some of which collected tags or annotations, others conducted situated crowdsourcing studies.
In the articles in which we could infer the number of tasks, the number ranged from 10 tasks \citep{2565585.2565608.pdf} to 55,000 \citep{18947-Article_Text-22713-1-2-20211004.pdf} (assuming one rating per task).
Because of the difficulty of determining the exact number, we do not report the mean and standard deviation of the number of tasks in these articles.

Monetary rewards were reported in 33~articles (\pt{33}\%) but often without explicitly mentioning MTurk fees. %If payments to workers were reported, it was often not clear if payment included the crowdsourcing platform fees or not.
% This adds to the opaqueness of the reporting.
%
Of the 34 articles, eight involved unpaid volunteers \citep{2502081.2502221.pdf,2971485.2971492.pdf,3328906.pdf,3349611.3355545.pdf,3209281.3209347.pdf,2596695.2596713.pdf,2470654.2470685.pdf,7472-Article_Text-10866-1-10-20200930.pdf},
two articles simply stated that participants were paid minimum wage \citep{2858036.2858321.pdf,2858036.2858327.pdf}, and
one article involved a raffle for an iPad \citep{2504776.2504796.pdf}.
Three articles reported the monetary rewards in the crowd pilot study as an average hourly \citep{13075-Article_Text-16592-1-2-20201228.pdf,3209542.3209558.pdf} or per minute wage \citep{2513383.2513448.pdf}.
Among the remaining 20 articles, the monetary pay for participating in the crowd pilot study ranged from \$0.01 \citep{2464464.2464482.pdf} to \$10 \citep{3479572.pdf} per task.
    \citet{2565585.2565608.pdf} paid \textit{``a maximum of \$30 US''} (p. 3), % \citep{2565585.2565608.pdf},
    but in this case, workers were recruited on Elance and ODesk (now Upwork) and the actual bids may have been lower.

Only few articles ($n=3$, \pt{3}\%) reported experimenting with different price points.
% This approach was taken in three articles. %  to find the optimal pay for the main study.
    \citet{2442576.2442591.pdf} experimented with paying workers in increments \textit{``from \$0.00 to \$8.00 [...] up to the federal minimum wage in the United States (\$7.25/hour as of April 2, 2011)''} (p. 4).
    Similarly, \citet{2897366.pdf} posted tasks \textit{``in \$.05 increments, starting at \$.15 and going up through \$.50''} (p. 10). % \citep{2897366.pdf}.
\citet{2903138.pdf} investigated the robustness of results against varying levels of reward (from US\$0.05 to US\$0.10). % \citep{2903138.pdf}.
Bonuses to workers in the context of crowd pilot studies, in general, were only mentioned in a few articles.
    \citet{2660114.2660126.pdf}, for instance, mention experimenting with different incentive schemes that involve a bonus to well-performing workers % \citep{2660114.2660126.pdf}
    and 
    \citet{2470654.2470743.pdf} conducted a crowd pilot study to determine a bonus based on the workers' average accuracy. % \citep{2470654.2470743.pdf}.

% The majority of articles ($n=116$, \pt{116}\%) did not report the number of workers participating in the crowd pilot study.
% not reported: 128 articles (\pt{128}\%)
% The vast majority of articles did not mention the pay to workers participating in the crowd pilot study ($n=138$, \pt{138}\%).

% --------------------
\subsection{Differences in how Crowd Pilot Studies are reported}%
% --------------------

% --------------------
\subsubsection{Are there differences in crowd pilot study reporting between research communities?}%
\label{sub:communities}%
% --------------------
%
As mentioned in Section~\ref{sec:corpus} and depicted in \autoref{fig:communities}, the bulk of crowd pilot studies were reported in three research communities: CHI, CSCW, and HCOMP.
The former two are closely related human-centered venues and researchers often submit to both venues. The latter is a venue specialized on advancing the state of the art of human computation and crowdsourcing, but also on applying it practically.
We explored differences between the three venues in how crowd pilot studies are being reported in human-centered conferences (CHI and CSCW) as compared to crowdsourcing research (HCOMP).
Our initial hypothesis is that there will be a difference between the communities since best practices will likely emerge from within the community of practice~\citep{Wenger2009CommunitiesOP} in the crowdsourcing-focused domain at HCOMP.

% -----
\begin{figure}[!h]%
\centering%
\includegraphics[width=.65\linewidth]{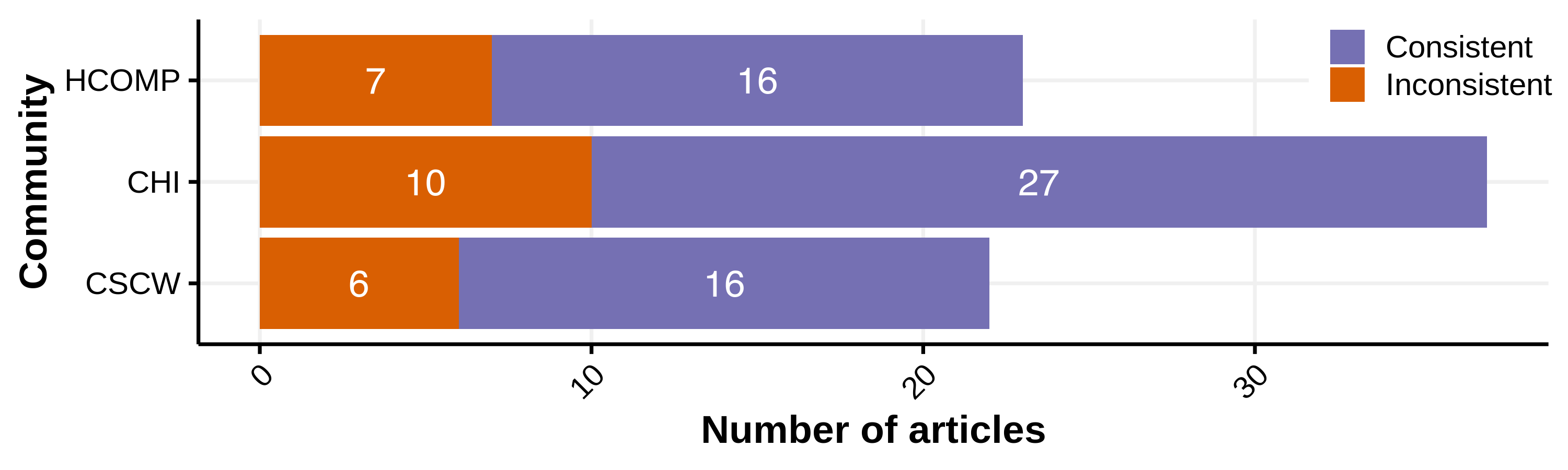}%
\caption{Consistency of wording within articles in different research communities.}%
    % \Description{Consistency of wording within articles in different research communities.}%
\label{fig:consistencycommunities}%
\end{figure}%
% -----

\FPeval\cchi{27/(10+27)*100}%
\FPeval\cchi{round(cchi:1)}%
\FPeval\ccscw{16/(6+16)*100}%
\FPeval\ccscw{round(ccscw:1)}%
\FPeval\chcomp{16/(7+16)*100}%
\FPeval\chcomp{round(chcomp:1)}%

\begin{figure}[!h]%
\centering%
\includegraphics[width=.8\linewidth]{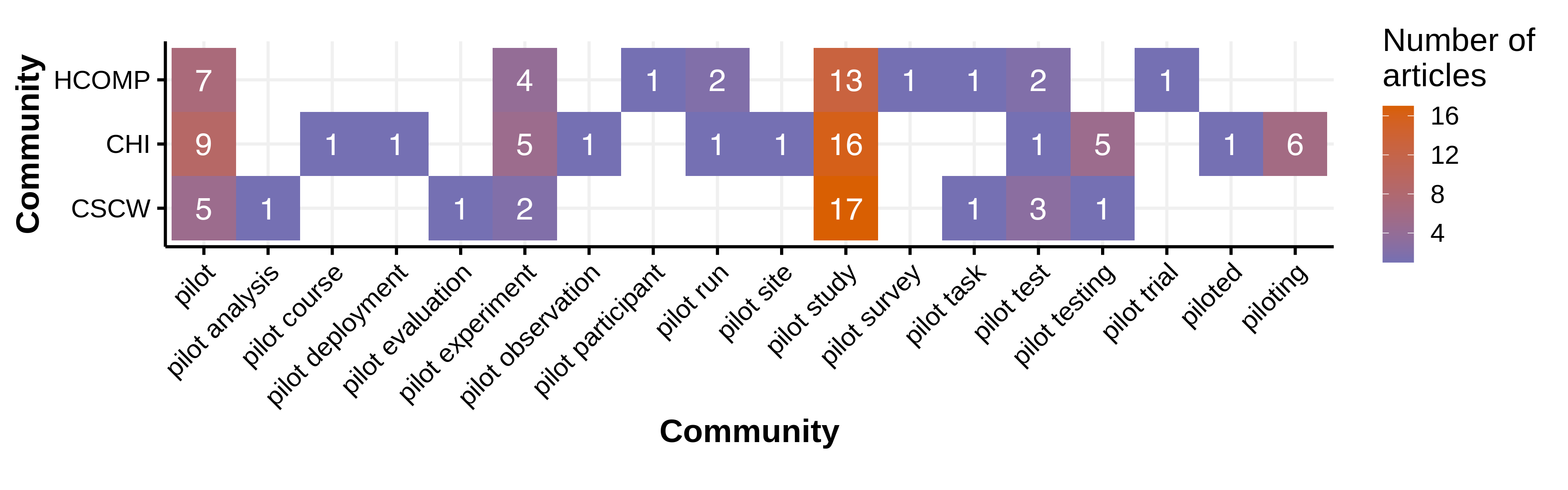}%
\caption{Wording used to refer to crowd pilot studies within articles in different research communities.}%
    % \Description{Wording used to refer to crowd pilot studies within articles in different research communities.}%
\label{fig:wordingcommunities}%
\end{figure}

We first investigate how consistent crowd pilot studies are being reported in the three communities.
We define `consistency' in the reporting of crowd pilot studies as the uniformity in the use of terminologies, methodologies, and presentation of results across the surveyed articles. Specifically, an article is deemed `internally consistent' if its descriptions, methodologies, and terminologies related to pilot studies remain coherent and unambiguous throughout the article's text.
In contrast, articles with varied references to pilot studies are deemed `inconsistent.'
Looking at the ratio of internally consistent articles in the three venues (cf. \autoref{fig:consistencycommunities}), we find CHI and CSCW are about comparable in consistency (\cchi\% and \ccscw\%, respectively).
The ratio of internally consistent articles published at HCOMP is slightly lower (\chcomp\%), but this difference is not significant (pairwise t-tests, each with $p>0.05$).
    % \st{That means within the HCOMP community, it is more likely that an article will use only one single term to refer to a crowd pilot study consistently throughout the article.}}
%
% However, this does not mean that there is consistency between articles and agreement within the communities.
We find there is no agreement between articles in the three venues on which term is used to denote pilot studies, even among the articles which use only one term.
A wide range of different terms are being used in the three communities, with `pilot study' being most common, followed by `pilot' and `pilot experiment' (see \autoref{fig:wordingcommunities}).

% -----
% -----

\begin{figure}[!h]%
\centering%
\includegraphics[width=.66\linewidth]{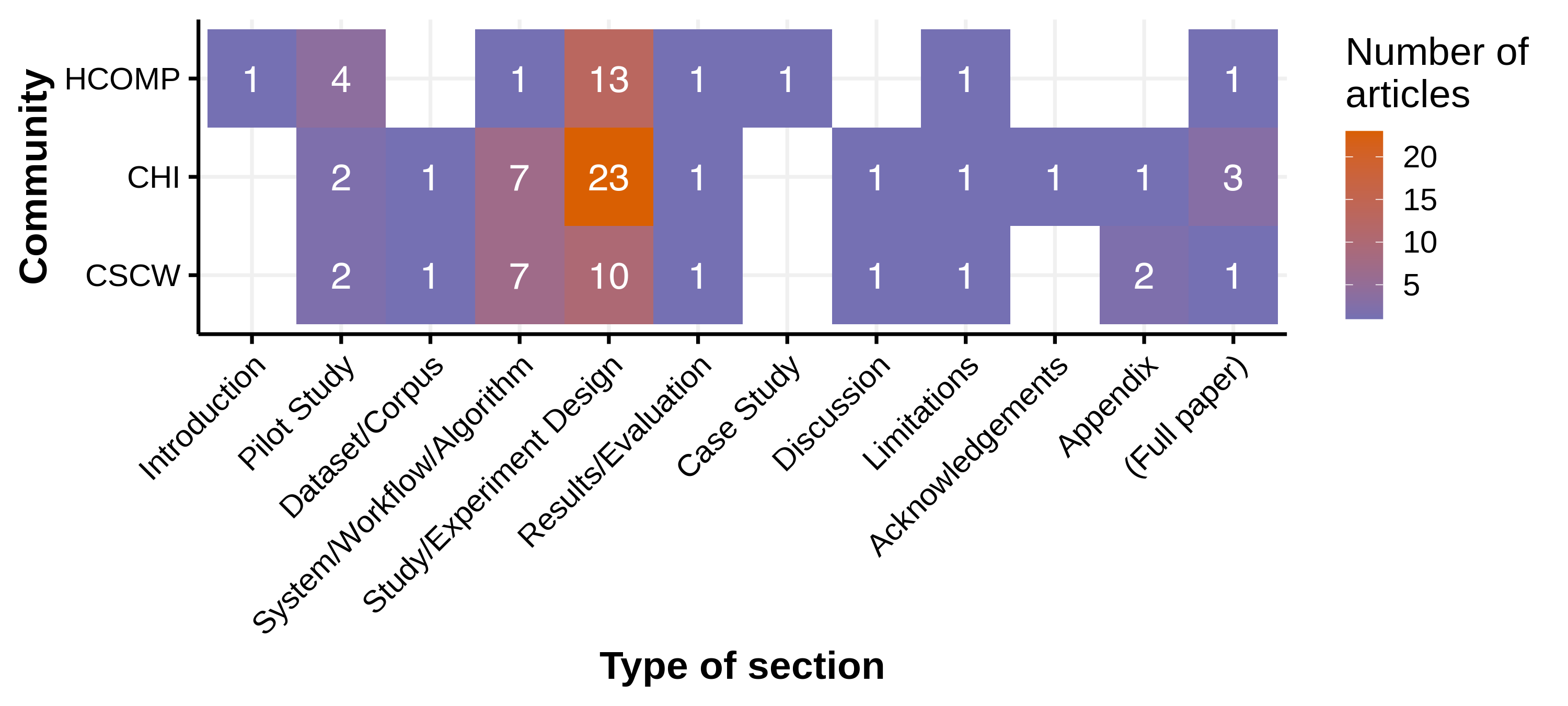}%
\caption{Type of section in which pilot studies are being reported in different research communities.}%
    % \Description{Type of section in which pilot studies are being reported in different research communities.}%
\label{fig:sectioncommunities}%
\end{figure}

Looking at the placement of pilot studies within articles (see \autoref{fig:sectioncommunities}), we find that crowd pilot studies are often reported in sections relating to the methodology (e.g., study design or experiment design).
There is no significant difference between the three venues when it comes to the section in which the pilot study is being reported, ${\chi}^2(22, N=90) = 19.37$, $p=0.6224$.
Four HCOMP articles (17.4\% of the articles in this venue) reported the pilot study in a separate section, which highlights the importance of pilot studies in the field of crowdsourcing, as compared to articles in CHI (5.4\%) and CSCW (9.1\%).
% -----
% -----

\begin{figure}[!h]%
\centering%
\includegraphics[width=.65\linewidth]{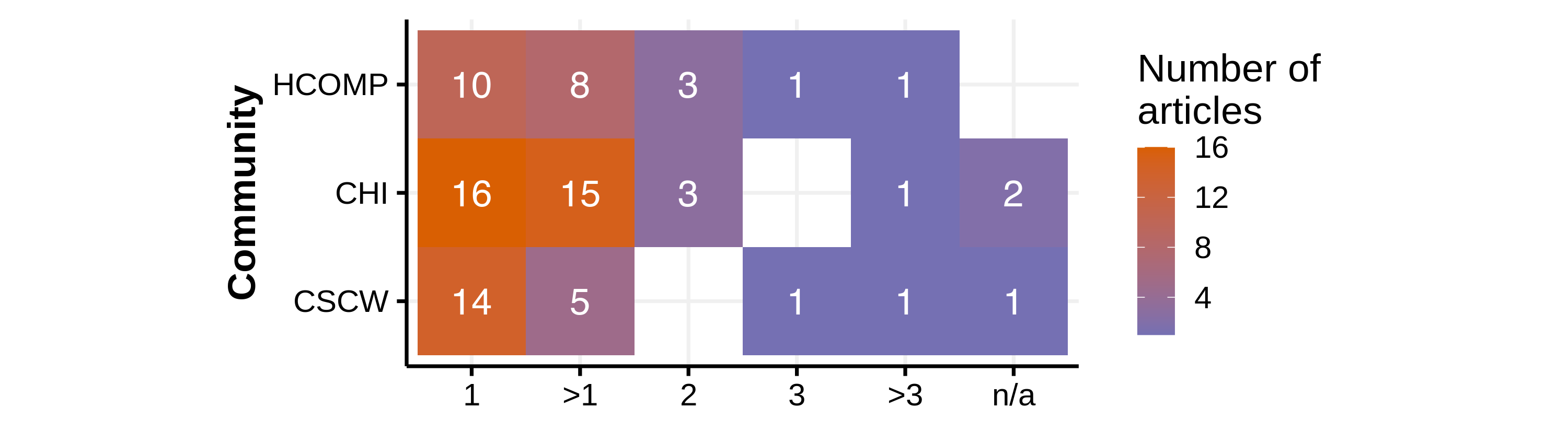}%
\caption{Number of crowd pilot studies conducted in the articles in different research communities.}%
    % \Description{Number of crowd pilot studies conducted per article in different research communities.}%
\label{fig:numpilotscommunities}%
\end{figure}

The number of pilot studies conducted within an article is similar in all three venues, with one single crowd pilot study being most common (see \autoref{fig:numpilotscommunities}).
At CHI and HCOMP, it is also common for articles to report more than one crowd pilot study.
%          No..of.pilot.studies
% Community   1   >1
%     CSCW   14    8
%     CHI    16   21
%     HCOMP  10   13
The difference between the three venues is, however, not statistically significant (${\chi}^2(10, N=82) = 8.3317$, $p=0.5965$).
% -----
% -----

\begin{figure}[!h]%
\centering%
\includegraphics[width=.8\linewidth]{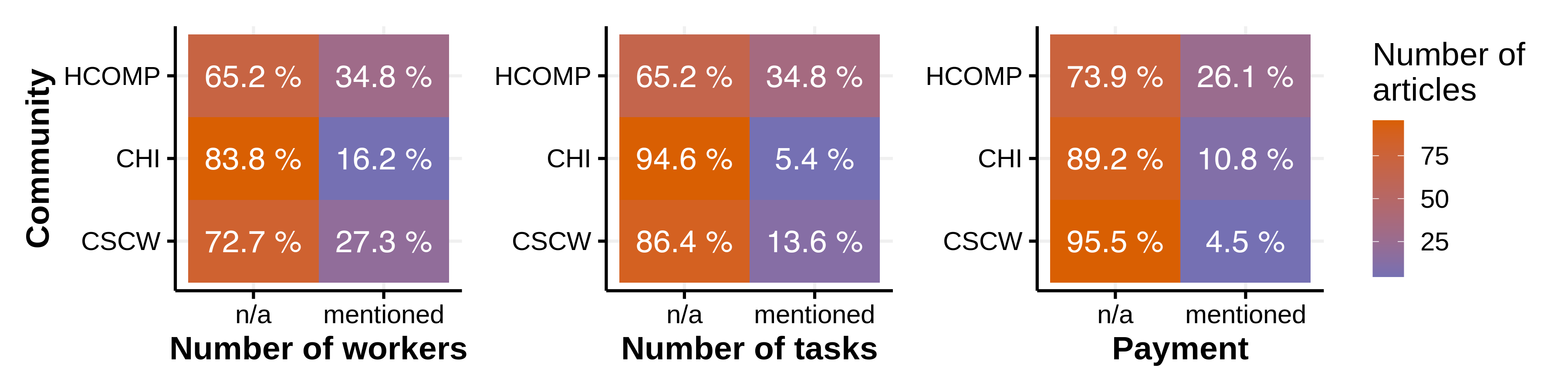}%
\caption{Key statistics reported in different research communities.}%
    % \Description{Key statistics reported in different research communities.}%
\label{fig:keystatscommunitiespercent}%
\end{figure}

When it comes to reporting key statistics about the crowd pilot studies, we find that in all three venues, the most common way of reporting a crowd pilot study is in passing without providing any details about the number of participating workers, the amount of tasks, or the exact monetary rewards provided to workers.
    A large percentage of articles (between 65\% to over 95\% of the articles reporting crowd pilot studies in each conference venue) do not report these three key statistics (see \autoref{fig:keystatscommunitiespercent}).
% -----
% -----

There are, however, differences between HCOMP and CHI/CSCW when it comes to reporting details about the crowd pilot study.
Authors in HCOMP are more likely to report key statistics about the crowdsourcing campaign as compared to CHI and CSCW (see \autoref{fig:keystatscommunitiespercent}).
% HCOMP articles are, compared to articles published in CSCW and CHI, more likely to report the number of workers participating in the crowd pilot study.
    HCOMP articles report the number of workers almost twice as often as CHI articles.
Similarly, the number of tasks assigned to workers is more likely to be reported in HCOMP articles (34.8\%) as compared to CHI (5.4\%) and CSCW articles (13.6\%).
This difference is even more profound when it comes to reporting payments to workers. HCOMP articles reported payments to workers participating in crowd pilot studies in about a quarter of the HCOMP articles as compared to CHI (8.1\% of the CHI articles) and CSCW (4.5\% of the CSCW articles).
One possible reason for this is that authors at HCOMP may be more sensitive to issues surrounding crowd work due to fairness of crowd work being a long-standing research topic in human computation and crowdsourcing. 
The differences between the three conference venues were, however, only statistically significant for the number of tasks, ${\chi}^2(2, N=82) = 9.2864$, $p=0.0096$.

% STATISTICAL DIFFERENCE ONLY FOR NUMBER OF TASKS.
% \\
% number of workers:
% X-squared = 2.7866, df = 2, p-value = 0.2482
% effect size (Cramer's V): 0.2607026
% \\
% number of tasks:
% X-squared = 9.2864, df = 2, p-value = 0.009627
% effect size (Cramer's V): 0.4759176
% \\
% payment:
% X-squared = 4.8858, df = 2, p-value = 0.08691
% effect size (Cramer's V): 0.3452041

In summary, we found no major % significant
differences between the HCOMP and CHI/CSCW communities in terms of the number of crowd pilot studies being conducted and the wording used within articles.
The consistency of wording within articles was comparable between the three venues, with many different terms being used to denote the crowd pilot study (some more common than others).
Authors in all three venues prefer to report the results of crowd pilot studies in a section relating to methods, with the study design section being the top choice of authors.
% Authors in HCOMP also reported the crowd pilot study in its own section, which speaks to the nature of the conference which seeks less to apply but to advance crowdsourcing.

% --------------------
\subsubsection{How do crowd pilot studies differ between academia and industry?}%
% --------------------
In our survey study, we asked participants if they had industry experience or worked closely with the industry. In response to this, four participants (36.36\%) responded they had industrial experience, while seven did not have experience (63.64\%). Two out of four (50\%) who had industrial experience do research in collaboration with an industrial partner, while one (25\%) indicated that he is planning to conduct a pilot study with industry. When asked what differences the participants found between the academic and industrial crowd pilot studies, one indicated that \textit{``industrial pilot studies cost more in salaries than academics''} while another respondent stated that \textit{``the pilot study was to create a digital asset for the company.''}

We also asked about potential differences between crowdsourcing crowd pilot studies on in-house/internal platforms and other commercial platforms (e.g., MTurk). Participants came up with a variety of feedback. One participant indicated that \textit{``the internal CS platform is more accurate than other commercial platforms''} and that \textit{``internal systems are easier to use as managers would have no issue with them, external ones are more tricky due to privacy and security issues.''}

% --------------------
\subsection{RQ4: What makes a ``good'' crowd pilot study?}%
\label{sec:RQ4}%
% --------------------
% pointers:
%  --What Features define a Crowd Pilot Study?
% -- what criteria need to be fulfilled to do a crowd pilot study?

In response to this question, researchers in our survey (cf. Section \ref{sec:survey}) identified several qualities that define a good crowd pilot study.
These qualities relate to the objectives for running a pilot study and may stand in tension with each other, as evident in the following sub-sections.

% ---
\subsubsection{Mimicking the main experiment}%
% ---

Two researchers stated that a successful crowd pilot study is \textit{``as similar to a formal experiment''} and \textit{``one that only differs from the complete study by sample size.''} This finding is also consistent with the recommendations of other researchers that a (crowd) pilot study should mirror all the processes of the main research and adhere to the identical protocol, including inclusion and exclusion criteria for participants, measuring tools, and training resources \citep{doi:10.4097/kjae.2017.70.6.601}.\par

% ---
\subsubsection{Exploration and experimentation}%
% ---

Others stated that the paramount quality of a good crowd pilot study is its exploratory nature, which provides them with different directions for their primary research questions or hypothesis. For instance, one noted that \textit{``[a good pilot is the] one which gives a clear direction of which RQs/directions would be more promising to pursue in an actual study''} and one that \textit{``should give researchers some useful inputs about their hypothesis or prototype.''} This finding shows that researchers use pilot studies in the conception phase of their projects when they need supporting evidence to develop a research question and research plan \citep{van2002importance}. \par

% ---
\subsubsection{Validating the feasibility of a study}%
% ---

Another quality mentioned by researchers is the ability of a crowd pilot study to assess the feasibility of an approach. This feasibility could also refer to the technical feasibility where researchers test rigorously through several trials that \textit{``all functions are working, and log [that the] system can repeat what users have done.''} Others viewed a good crowd pilot study as one that \textit{``allows to validate the functioning of your task and it allows you to gather a sample of the final expected data''} because \textit{``it costs high to redo a formal exp.''} Other respondents defined assessing the task related information as the main criteria that a crowd pilot study should incorporate, such as the \textit{``task length, number of workers required, task complexity and task design''}. For instance, one researcher summarized this in the following words \textit{``I think we need to always do a pilot study, to figure out if both the design and the technical problems are solved.''}

% ---
\subsubsection{Accurate estimation of campaign parameters}
% ---

Another critical dimension that defines a good crowd pilot study is its ability to estimate sample size and power calculations. For instance, researchers reported that a good crowd pilot study could help to \textit{``[correct] the sample size errors''} and \textit{``help to calibrate the power calculation.''} Similarly, one participant reported that a crowd pilot study should help to estimate the \textit{``statistical data involved, e.g., mean/median/sd.''} Thus, sample size and power calculations are another essential quality of a good crowd pilot study. These estimates are even crucial in crowdsourcing research when researchers need to hire hundreds of participants, thanks to the affordability and affordances of crowdsourcing platforms. However, the estimation of the sample size required for the main trial needs to be performed cautiously since a crowd pilot study only provides the estimated value of standardized effect size \citep{LEON2011626}. Moreover, one may also need to account for participant exclusion in such cases.

% % --------------------
% \subsubsection{How does the Crowd Pilot Study differ from the Main Study?}%
% % --------------------
% %

% How does the crowd pilot study differ from the main study in terms of worker compensation, cost ratio, platform capabilities, and institution--industry ties?

% --------------------
% \subsection{RQ6: What are the Impediments to reporting Crowd Pilot Studies?
% What are the Factors that promote or obstruct reporting Crowd Pilot Studies?}%

\subsection{RQ5: What factors promote or obstruct the reporting of crowd pilot studies?}
\label{sec:RQ5}%
\begin{figure}[!h]%
\centering%
\includegraphics[width=.6\textwidth]{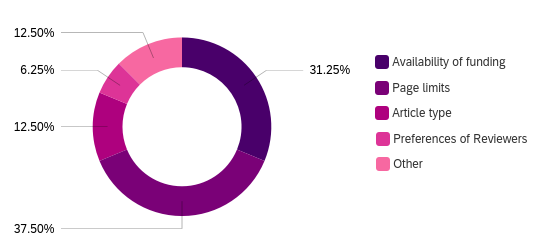}%
\caption{Factors that promote or inhibit the reporting of crowd pilot studies.}%
    % \Description{}%
\label{fig:RQ5}%
\end{figure}

% --------------------
We approached this question by posing both closed- and open-ended questions.
In a closed-ended question, we asked participants to provide potential reasons for characteristics that either encourage or restrict the reporting of crowd pilot studies.
These reasons include page limit restrictions, funding availability, article types, and reviewer preferences (see \autoref{fig:RQ5}).
% The results are shown in \autoref{fig:RQ5}.
% The second part of this question was open-ended and asked respondents to elaborate % on why they believed the factor they selected was the reason.
% and justify their answers.
Most participants indicated that page limitation were the most critical factor ($n=6$), followed by availability of funding ($n=5$). Only two respondents answered that the article type was the most essential factor, and one responded that he does crowd pilot studies because reviewers want them.%

% ---
\subsubsection{Page limit restrictions}
% ---
Regarding the \textit{page limitations}, respondents felt that \textit{``it limits the content length of the report''} and they would prefer actual experiments over crowd pilot studies \textit{``because the results of the formal experiment are more interesting [than pilot studies].''} Another respondent believed that \textit{``conf[erence] papers normally require a tight page limit which would squeeze the space for rather important content (e.g., results).''} % if reported pilot.''}
Another respondent who worked in the area of crowd-powered applications responded that \textit{``justifying some design decisions of a big crowd-powered system is probably not very critical. We will likely cut these justifications when we don't have space.''} We also asked ``if there is no page limit, will this make you more likely to report crowd pilot studies in your articles?''

% The results can be seen in \autoref{fig:RQ5}.
Three out of eleven respondents believed that they will `very likely' report pilot studies, two affirmed that they would undoubtedly report pilot studies, while one was neutral about this opinion. We also noted that no respondents selected `unlikely' or `highly unlikely', which shows that page limitation is a rather decisive factor. This trend is slowly shifting. For instance, conferences have been slowly transitioning to a revise and resubmit cycle along with more flexible manuscript lengths, which removes page restrictions and permits authors to expand the methodology and design sections, enabling the reporting of crowd pilot studies.%

% ---
\subsubsection{Availability of funding}
% ---
Participants also felt that the \textit{availability of funding} may encourage the scalability of an experiment and extensive testing of a product before it could be made available.
A participant, for example, % phrased this as:%
thought that
% \begin{quote}%
    \textit{``funding is crucial in scaling the experiment, and the funding sources tend to encourage folks to include pilot results in the grant application.''}
% \end{quote}
Another participant who believed crowd pilot studies were important for iteration and testing stated this as follows:
\begin{quote}
      \textit{A good project needs to be developed and tested for a long time before it can be released. Therefore, the initial investment is relatively large and stable sources of funds are needed.}
\end{quote}

% ---
\subsubsection{Article types}
% ---

The \textit{article type} also played a significant role in inhibiting the reporting of a pilot study. For example, one respondent responded as:%
\begin{quote}
    \textit{To report this we need to write a paragraph or at least some sentences describing it. If we need to cut down something due to exceeding the page limit, this would be an option. For conference or journal, because the target audience have different focus. For some system-focused venue, we may shorten the description of data collection and experiment design by skipping this.}
\end{quote}

% ---
\subsubsection{Reviewer preferences}
% ---

One respondent was of the opinion that \textit{``reviewers perceive pilot studies as less impactful and therefore would not be willing to accept them for publication.''}

% \begin{figure}[!h]%
% \centering%
% \includegraphics[width=.65\linewidth]{figures/images/survey/RQ5_PART2.png}%
% \caption{This figure demonstrates the percentage of respondents who indicated they would report pilot studies if there were no page constraints.}%
%     % \Description{This figure demonstrates the percentage of respondents who indicated they would report pilot studies if there were no page constraints.}%
% \label{fig:RQ5}%
% \end{figure}%

% --------------------
\subsection{RQ6: How can crowd pilot studies be facilitated with platform-specific features?}%
% \subsection{Implications for the Design of Crowdsourcing Platforms}%
% \subsection{Affordances of Crowdsourcing Platforms Facilitating Pilot Studies}%
\label{sec:RQ6}%
\label{sec:designimplications}%
\label{sec:affordances}%
\label{sec:features}%
% --------------------
%

In our literature review, we found a handful of mentions of platform-specific technical features that were used for conducting and monitoring crowd pilot studies.
In the remainder of this section, we reflect on the design of such features, based on our literature review, the results from our survey study, and our experience with different crowdsourcing platforms.

% ---
\subsubsection{Exit surveys for facilitating crowd pilot studies}
% ---
One possible feature for supporting crowd pilot studies is the `exit survey.'
    An exit survey is a short questionnaire that workers fill out after completing tasks.
Exit surveys were used by a few authors to measure or monitor workers' satisfaction with the payment during crowd pilot studies.
    For instance, \citet{2818048.2819979.pdf} used the aggregated results of the exit survey (provided by CrowdFlower/Figure Eight) to validate and justify the choice of payment. The authors reported the results of the exit survey indicated \textit{``that the chosen payment was considered acceptable by the workers''} and that \textit{``the selected compensation was appropriate for the specific study setting''} (p. 265). % \citep{2818048.2819979.pdf}.
Another use for an exit survey is collecting demographics, which is especially important on microtask platforms where tasks are typically too short to collect demographics.
    For instance, \citet{3230665.pdf} used a custom exit survey on Amazon Mechanical Turk to collect demographic information. % \citep{3230665.pdf}.

% ---
\subsubsection{Reward calculation}
% ---

Besides the above feature, participants in our survey mentioned a number of other features that could facilitate crowd pilot studies.
% reward calculator
Most often mentioned was a ``reward calculator'' which could calculate rewards based on estimated completion times. As found in our literature review, the calculation of rewards from average task completion times is one of the most common reasons for conducting crowd pilot studies.
Prolific\footnote{\url{https://www.prolific.co}}, a crowdsourcing platform for academic studies, already offers a recommendation for the price of a task, based on the estimated time.
This is, however, only an incomplete solution because it is difficult for a requester to estimate the completion time~--- often, the very reason for conducting the crowd pilot study is finding this estimate.
However, crowdsourcing platforms are host to many different types of tasks. Given the large variety and amount of tasks on the crowdsourcing platforms, it would be possible for platform operators to collect information on tasks and to devise machine learning based platform features to support the estimation of task completion times and task rewards, based on empirical data collected on the crowdsourcing platform.

% ---
\subsubsection{Better support for running qualification studies}%
% ---
%
Screening criteria were mentioned often by the survey participants.
% Crowdsourcing platforms vary in their support for screening and qualification studies.
Crowdsourcing platforms differ in their capabilities to support screening and qualification studies.
Custom qualifications can be created, but this requires running a study, collecting results, and then uploading a comma-separated values (CSV) file to Amazon Mechanical Turk to assign the custom qualifications to workers. Only then can the qualification be selected in future studies.
Amazon Mechanical Turk offers only a limited set of qualification criteria for screening participants.
% While qualification studies on Prolific are supported by a more extensive and robust list of pre-defined qualification criteria, custom qualification studies are also more complicated (if implemented with a survey study) or limited to Prolific's built-in multiple choice survey studies).
Although Prolific offers a broader array of pre-defined qualification criteria, setting up custom qualification studies can be just as complex as in MTurk (when implemented via a survey study) or restricted to Prolific's in-built multiple-choice survey options.
Better user interfaces for running qualification studies and setting (or deleting) qualifications are needed.
% extending a study with more places
The survey participants further perceived a need for an MTurk feature to extend running studies with more participants.
On Amazon Mechanical Turk, no changes can be made to a running crowdsourcing campaign. This leads to disparate sets of survey results which then need to be manually integrated by the researcher.
% communication with workers
Last, the participants in our survey mentioned wanting better tools to communicate with workers, such as a chat or e-mail service.
This speaks to the survey participants' need for a less dehuminizing communication with crowd workers~\citep{chi_rehumanized_crowdsourcing.pdf}.
Features to communicate with the crowd would allow requesters to better monitor ongoing studies and grow a base of trusting participants \citep{doi:10.1080/08956308.2017.1325689}.

%%% Amazon Mechanical Turk has a sandbox for internal testing of tasks.
%%% But sandbox is only for internal pilot testing.
%%% Maybe there should be a separate sandbox only for pilot tests?

Clearly, there is an opportunity for the design of dedicated features on crowdsourcing platforms that could better support and facilitate running crowd pilot studies.
Features, such as the above, could
% prove useful for both requesters and crowd workers and 
support best practices in crowdsourcing.
% <TRANSITION>
We reflect on the importance of best practices and make recommendations for reporting crowd pilot studies in the following section.

% ====================
\section{Moving forward: Best practices for reporting crowd pilot studies}%
\label{sec:discussion}%
% ====================
%
Crowd pilot studies are a common and required method in crowdsourcing research due to the empirical nature of the crowdsourcing paradigm.
Unsurprisingly, many authors report having conducted crowd pilot studies in the scholarly literature. Yet, no scientific study spanning crowdsourcing has investigated this topic in depth. Our work aimed to fill this gap.
In this section, we reflect on our findings and the current state of best practice on reporting crowd pilot studies.

% < need to relate all this to CHI / HCI ??? takeaways for HCI / CHI ?? >
% <need to better relate all this to related literature>

% <TRANSITION>

% --------------------
\subsection{Readdressing current practices for reporting crowd pilot studies in crowdsourcing research}%
\label{sub:relationtoHCI}%
% --------------------
%
Crowd pilot studies connect to two strains of research in the field of crowdsourcing that % are seated at
touch upon the very nature of crowdsourcing: 
fair and responsible crowdsourcing \citep{5283-Article_Text-8384-1-10-20190926.pdf,p39-silberman.pdf}
as well as reproducibility in empirical computer science \citep{3360311.pdf}.
These two issues have long been debated in the scholarly literature.

% --------------------
\subsubsection{Best practices for fair and responsible crowdsourcing research}%
% --------------------
Crowd pilot studies account for a significant amount of work that is unaccounted for to a large extent in the scholarly literature.
Since the majority of authors use opaque language masking the extent of studies, little is known about the real extent of crowd pilot studies.
Further, due to the empirical nature of crowdsourcing, it is likely that crowd pilot studies often underpay participants. Estimating the rewards for crowdsourced tasks is hard and one way of adressing this shortcoming is to raise the basic level of payment or assign bonuses to workers in a post-hoc manner to fairly compensate participants in crowd pilot studies~\citep{qiu2021using,balayn2022ready,he2022}. However,  it is likely that workers in crowd pilot studies are substantially and potentially systematically underpaid~\citep{data-driven-analysis-of-workers-earnings-on-amazon-mechanical-turk.pdf,f5caef4f46d2a37d210b9e9811cb207921ba.pdf,wsdmf074-difallahA.pdf}. Recent work has unearthed different forms of invisible labor that crowd workers put in as they strive to earn their livelihood in various crowdsourcing marketplaces~\citep{3476060.pdf,ghost}. Prior work has also revealed how crowd workers are often subject to unfair rejections following qualification studies~\citep{mcinnis2016taking,gadiraju2019understanding,edixhoven2021improving}. It is likely that such practices transcend to ill-reported crowd pilot studies.  
Interestingly, extremely few articles in our literature review reported that bonuses were given to workers in or after crowd pilot studies. Based on results from our literature review and survey, we find it is more typical~-- though still not common~-- to pay bonuses to participants for performing well in the main study.

% The power dynamics on crowdsourcing platforms are in the favor of requesters and crowd workers have long been termed an invisible work force \citep{3476060.pdf,f5caef4f46d2a37d210b9e9811cb207921ba.pdf} or ghost workers \citep{ghost}.

% Crowd workers on paid microtasking platforms are an invisible labor force \citep{3476060.pdf}.
% % The power dynamics on crowdsourcing platforms are in favor of requesters.
% \citeauthor{p39-silberman.pdf} urged requesters to ``pay crowdworkers at least minimum wage'' \citep{p39-silberman.pdf} and \citeauthor{5283-Article_Text-8384-1-10-20190926.pdf} developed a script to raise the pay to workers to a minimum level \citep{5283-Article_Text-8384-1-10-20190926.pdf}.

% --------------------
\subsubsection{Reproducibility in crowdsourcing research}%
% --------------------
The crowdsourcing paradigm has many known limitations.
For instance, results obtained from crowdsourcing studies may be difficult to reproduce due to the anonymity of the workforce.
    % and results typically do not generalize to larger populations.
The opaque reporting of crowd pilot studies, as evidenced in our literature review, adds one additional layer to the issue of reproducibility. The strong prevalence of reporting on study results in passing accentuates and entrenches bias in research and helps bad practices to endure.
For instance, some authors used the crowd pilot study to substantiate claims. Crowd pilot studies are sometimes used as a magic linguistic device to materialize results that are later used as input for the main study of the article. In this sense, much of the reporting on crowd pilot studies uses `hedging' language, a ``rhetorical means of gaining acceptance of claims''~\citep{hyland1996.pdf}.

We argue that researchers should do their due diligence on the research claims made and report transparently on the aims and results of crowd pilot studies.
Readers need to know the details about crowd pilot studies.
    For instance, a reader needs to know the number of participants in a crowd pilot study ``to know that the study was big enough to justify the claims made'' \citep{publomics}.
Authors need to realize that opaque reporting on crowd pilot studies --- especially if it is done as a summative evaluation (as was the case in a few articles we reviewed) --- weakens the claims of the authors' research.
% Further, not reporting on pilot studies may lead to biased results.
Insufficient details can impede the progress of science in general.
The current state of reporting on crowd pilot studies exacerbates and affirms this widespread practice of opaque reporting.
More transparency on reporting crowd pilot studies is needed to nudge the current state of reporting in the field of crowdsourcing toward a code of practiced ethics that values transparent reporting of crowd pilot studies.
% However, as we discuss later in this work, this is not an easy task and may require a systemic shift.
%
% UG: Add references, draw a parallel to crowd work, practices
However, crowdsourcing is still a relatively young field where good practices need building.

\subsubsection{Treating the crowd workforce fairly.} One of the pivotal realizations that has emerged through research and practice within the crowdsourcing community over the last few years is the need to treat crowd workers fairly and with dignity -- whether it is in terms of the hourly wages paid or with respect to communication with workers~\citep{5283-Article_Text-8384-1-10-20190926.pdf,turkopticon.pdf,p39-silberman.pdf}. It is now commonplace in most HCI communities to declare the hourly wage that participants are paid in reported main studies. By raising the bar for what is expected in the reporting of crowd pilot studies in scholarly literature, we can hope to instill the otherwise (potentially) dormant desire to pay workers fairly within crowd pilot studies. This will increase the overall accountability of researchers and other requesters, and help bridge a gap in the invisible labor prevalent in crowdsourcing marketplaces~\citep{ghost}. 

Beyond crowd pilot studies, the broader domain of data annotation stands as another significant area where fairness in treatment and payment of crowd workers is paramount. Data annotators play a foundational role in shaping machine learning models and other AI systems by providing high-quality labeled data. Yet, there have been growing concerns about the remuneration, working conditions, and well-being of these data annotators, especially given the labor-intensive nature of their tasks \cite{turkopticon.pdf,3411764.3445092.pdf,CHI2020workshop,ghost,3491102.3501866.pdf,3491102.3502121.pdf}.
Inadequate compensation for data annotators not only poses ethical dilemmas but also risks compromising the quality of annotated datasets. By ensuring fair wages and conditions for these workers, we not only uphold the principles of ethical research and practice but also contribute to the production of more reliable and robust AI systems. It is crucial for the HCI and broader AI communities to address this concern head-on, establishing standards that reflect the true value of this indispensable labor.

%UG: I will write a paragraph here about how reporting pilot studies in detail can also bring about a change in how researchers treat payment in pilot studies.

%<TRANSITION (to section 5.2):>
Creating a widespread change in how crowd pilot studies are reported will require widespread and collective action. This is especially required, since well-meaning authors are often subject to a trade-off while reporting crowd pilot studies, as discussed in the following section.

% --------------------
\subsection{The trade-offs around reporting crowd pilot studies}%
\label{sub:tradeoffs}%
% --------------------
Researchers are influenced~-- consciously or unconsciously~-- in how they report crowd pilot studies.
In this section, we % summarize and
discuss confounding factors and biases that may affect the reporting of crowd pilot studies in academia and industry.

\subsubsection{Page limitations}
Traditionally, the page limit at conference venues such as CHI was 10 pages (in two-column format, not including references). % This explains the spike in our dataset around that number of pages (cf. \autoref{fig:pages}).
Some venues continue to uphold such strict restrictions on the number of pages in articles.
Authors, therefore, face a difficult trade-off between reporting on pilot studies in detail and reporting on the main study.
Our literature review is evidence for this trade-off, with many crowd pilot studies being reported briefly and casually.
In recent years, however, many venues in HCI (e.g., CHI and CSCW) have relaxed the page limitations which could, in theory, encourage authors to dedicate more space to crowd pilot studies.
However, a number of other biases and trade-offs may still make authors consider otherwise, such as the academic publishing model.

\subsubsection{Academic publishing model}
Publication bias is ``the failure  to  publish  the  results  of  a  study  `on  the basis  of  the  direction  or  strength  of  the  study findings'{''} \citep{53.full.pdf}.
Due to the publication bias in academia, authors may feel the need to report positive findings in order to get published.
A formative crowd pilot study, in particular, may -- in the mind of authors and/or reviewers -- not add to this goal.
Further, if authors feel the pilot study does not contribute towards the acceptance of the article, the authors may decide to omit the pilot study or shorten the reporting.
Another concern that authors may have is that if a formative pilot study is given too much space in the article, reviewers may view the article as a work-in-progress and recommend it for acceptance in a lesser capacity (e.g., as a poster).
Therefore, researchers may decide not to report pilot studies because of a perceived need to produce writing that pleases reviewers.
However, in the past years, some conference venues have opened up to the possibility of submitting works following the principles of open science. These venues encourage the submission of replications and articles with null or negative results which have been traditionally hard to publish.
While these advances, so far, have been limited to special tracks~-- such as the Open Science track at the Conference on Intelligent User Interfaces (IUI) 2023~-- they could lead to a slow systemic change toward an academic system in which reporting on pilot studies is being encouraged.
In this regard, some referees may consider it favorable if crowd pilot studies are being reported transparently and in detail.

\subsubsection{Funding}
The availability of funding is another important factor that may influence the authors' decision to conduct or report pilot studies.
For instance, the availability of funding may affect the extent of crowd pilot studies.
    If researchers are short on budget, they may skip or reduce the number of formative pilot studies.
However, even if funding is available, authors may decide not to run crowd pilot studies in order to not ``waste'' the funding organization's money on formative studies with an anonymous crowd.
For similar reasons, authors may decide not to report on crowd pilot studies.
On the other hand, iterative experimentation is important, especially in the field of Human-Computer Interaction (HCI) where emphasis is placed on iterative and participatory design to ensure optimal outcomes in a variety of contexts~\citep{benyon2013designing}.
The very process of design requires iteration and formative experimentation to arrive at an acceptable solution.
% <something more to conclude this section>

\subsubsection{Corporate or organizational culture}
If not the academic system or external funding, then the internal culture of an authors' organization could discourage conducting pilot studies.
For instance, universities in Finland recently started following stricter directives from the Tax Administration in a move towards a system where any rewards to participants~-- whether it is cinema vouchers, gift cards, or monetary payments~-- need to be declared to the tax office, regardless of the monetary value of the rewards.
Because monetary compensations to participants are subject to withhold tax, this causes an overhead to the university administration.
Even more concerning is that researchers are asked to collect private information from their study participants (name, address, and social security number), if participants are to be rewarded. %%% a GDPR nightmare
Therefore, researchers in Finland are strongly discouraged from using paid participant samples in their research.
This development is deeply worrisome as it discourages researchers in Finland from conducting ethical and fair science.

% --------------------
\subsection{Guidelines for Reporting Pilot Studies}%
\label{sec:guidelines}%
% --------------------
%
Our analysis of the HCI and crowdsourcing literature allows us to provide recommendations for reporting crowd pilot studies.
In this section, we revisit the research questions RQ1--RQ3 and
connect the findings of our literature review to recommendations
    % on the basis of what has been found to be the most common way of
for reporting crowd pilot studies.%
\subsubsection{Why are crowd pilot studies typically conducted? (RQ1)}

\paragraph{Be transparent on the reasons for conducting crowd pilot studies}%
Most articles in our literature review conducted crowd pilot studies for formative reasons. However, in articles that report crowd pilot studies in passing, it was sometimes not explicitly stated why a pilot study was conducted.
Clearly motivating the crowd pilot study will provide clarity to the writing and increase the readers' understanding of why a pilot study was needed.

\paragraph{Inform crowd workers that they are participating in a crowd pilot study}%
Crowd workers are subject to a wide range of tasks posted on crowdsourcing platforms. Some tasks are more and some less lucrative for the workers. Crowd pilot studies may fall into the latter category, especially if the task price is not estimated accurately.
While some workers are not motivated by extrinsic factors and may enjoy participating in crowd pilot studies \cite{crowdsourcingcreativity}, other workers may want to avoid them.
Workers should be informed that they are participating in a small-scale study (that may potentially be under-priced).

\subsubsection{How are crowd pilot studies typically reported? (RQ2)}%
\paragraph{Use consistent wording}%
Academic writing requires precise language. Using different terms within an article to refer to crowd pilot studies may add confusion to an uninitiated reader.
We recommend to use the term `pilot study' to denote crowd pilot studies. This term was the most common term used in the literature (cf. \autoref{fig:wording}).%
\paragraph{Report crowd pilot study findings in one section}%
% Some authors in our corpus reported findings from the crowd pilot study in multiple sections of their article.
We found many authors scatter findings from their pilot studies throughout their papers. To improve the clarity of pilot study reporting and to better showcase the results of pilot studies, 
% "In the guideline, simply suggesting “reporting findings in one section” does not make sense to me; also, simply indicating the most common practice is to report the findings in design-related sections does not provide a clear direction of what is the best practice of doing so. Same as the “consistent word” guideline, I am not sure whether it is an important to advice future researchers to follow (i.e., use “pilot study” rather than other variations) because different language usage might have different purposes/intentions behind it."
we recommend to bundle the reporting of crowd pilot studies in a single section of the article.
This would improve both the understanding of the reader of the extent of pilot studies conducted, and the reproducibility of the pilot study.
% Ideally, this would be a whole section dedicated to the results of the crowd pilot study, but page limitations may necessitate reporting findings from crowd pilot studies in other sections.
Our analysis of the literature indicates that it is most common to report the results of crowd pilot studies in design-related sections (cf. \autoref{fig:section}).%
\subsubsection{What do crowd pilot studies report? (RQ3)}

\paragraph{Report the number and extent of crowd pilot studies}%
A considerable amount of articles in our literature corpus (62 articles, about 36\%) did not provide information on the exact number of crowd pilot studies being conducted. In other cases, the number of studies being conducted had to be calculated from information scattered in the article.
Authors should clearly state the number of pilot studies and their respective extent.

\paragraph{Clearly identify the participants}%
There should be no room for interpretation when it comes to who participated in the crowd pilot study. In particular, researchers should identify whether crowd workers participated in the pilot study or whether the pilot study was conducted with a different participant sample (e.g., students, experts, or internal participants).
This also includes information on the crowdsourcing platform used in the pilot study, if it cannot be reliably inferred from the context in the article.
Internal pilot studies should be clearly denoted as such.

\paragraph{Report the key attributes of each crowd pilot study}%
If page restrictions limit authors from reporting in-depth on crowd pilot studies, we recommend to include at least the following key information when reporting crowd pilot studies:
\begin{itemize}%
    \item number of participating crowd workers,
    \item number of tasks (or assignments to workers),
    \item payment per task,
    \item participation constraints enforced (including platform settings), and
    \item the type of crowd or crowdsourcing platform.
\end{itemize}
The latter could be omitted if it is clear from the context that only one crowdsourcing platform was used throughout the article.
The selection of rewards to workers in the crowd pilot study should be justified. If there are major discrepancies between the rewards paid in the crowd pilot study and the main study, it should be explained how these discrepancies came into existence and what measures were taken to remedy the discrepancies.

% --------------------
\paragraph{Report a minimum set of information}%
\label{sec:minimumset}%
% --------------------
%
Inspired by scientific reporting guidelines, such as guidelines by the \citet{APA}, and based on the above recommendations while also considering the trade-offs discussed in Section~\ref{sub:tradeoffs}, we propose a condensed format for reporting formative crowd pilot studies:

\begin{framed}%
\centering%
    \textit{\ldots pilot study (MTurk; N=12; 1000 HITs; US\$4.5 per HIT) \ldots}
\end{framed}%

In combination with any of the preferred methods of reporting on pilot studies, such as
    \textit{``in a pilot study (\ldots), we found\ldots''}
    or
    \textit{``{\ldots}based on a pilot study (\ldots)''}
    (cf. \autoref{fig:phrasing}),
this condensed format provides key statistics about the formative crowd pilot study (i.e., the crowdsourcing platform, number of participants, number of HITs, currency, and price per HIT) without taking up an undue amount of space in the article.

We hope that authors will adopt at least this condensed way of reporting crowd pilot studies to increase the transparency and reproducibility of their research.
In the same vein, we hope that reviewers in conferences and journals publishing research with crowd pilot studies will, in the future, place increased emphasis on seeing transparent reporting of crowd pilot studies.

% \subsubsection{Optionally report other statistics relevant to the main study}%

% \subsubsection{What makes a ``good'' crowd pilot study? (RQ4)}
% \subsubsection{What are the factors that promote or obstruct reporting crowd pilot studies? (RQ5)}
% \subsubsection{How can crowd pilot studies be facilitated with platform-specific features? (RQ6)}

\subsection{Practical Suggestions for Supporting Better Crowd Pilot Study Reporting}%

% \todo{
% Authors' adoption of best practices could be facilitated with dedicated platforms for reporting crowd pilot studies.
% Such platforms could eliminate or lessen the factors that obstruct the reporting of crowd pilot studies.

% One suggestion to encourage proper pilot study reporting
% could be the creation of a repository of crowdsourcing pilot studies.

% [with their recommended reporting style] that authors
% can properly cite in their papers, so authors don't have to exceed page limits.
% }

Creating a centralized repository for crowd pilot studies in crowdsourcing research could be a possible way to enhance the reporting and transparency of such investigations. The repository would serve as a dedicated platform for researchers to submit their crowd pilot studies following a standardized report format, as suggested in the previous section.
    This format should encapsulate critical elements including research questions, methods, results, and challenges encountered, thereby enabling a comprehensive understanding of the study without exceeding paper page limits.
    Furthermore, the structured reporting style within the repository should include specifics such as sample size, study duration, data cleaning methods, and outcomes.
    This standardized approach, coupled with a requirement for authors to detail challenges and potential improvements, could not only foster transparency but also provide insights for researchers undertaking similar studies.

To further encourage crowd pilot study reporting, a system of incentives could be introduced for authors who make use of the repository, ranging from formal acknowledgments within the academic community, citation opportunities, to reduced publication fees in affiliated journals. At the same time, scientific journals and conferences could set forth clear guidelines encouraging the citation of crowd pilot studies from the repository in their submissions.
Creating this repository and encouraging its use would help cultivate a research culture that values transparently reporting crowd pilot studies, ultimately leading to more accurate, rigorous, and replicable crowdsourcing research.

% ====================
\subsection{Limitations}% and considerations}%
\label{sec:limitations}%
% ====================
%
%%% REPRESENTATIVENESS
In our literature review, we made pragmatic choices to limit the set of literature to what we believe is a representative coverage of the literature, as mentioned in Section 
\ref{sec:creatingcorpus}.
Our screened corpus included all articles from HCOMP, the premier venue for crowdsourcing research. The ACM Digital Library contains articles from human-centered journals and conferences, such as CHI and CSCW.
We do not claim that the selected corpus generalizes to all publications involving pilot studies.
However, this corpus provided a good view into the prevailing practices in diverse research communities.
     
%%% CHOICE OF KEYWORD
Another aspect in achieving representativeness is the choice of search keyword.
Our literature review may have missed articles that do not contain the `pilot' keyword and, instead, refer to the pilot study in other ways, such as ``preliminary study'' or ``formative study.''
However, both our survey and scoping review of the literature found that `pilot study' is the most common term to refer to crowd pilot studies.
Further, there is a difference between pilot studies and preliminary studies. The latter are primarily conducted to identify user needs and for defining requirements. In the context of crowdsourcing, however, pilot studies are being conducted for the specific purpose of determining and validating the parameters of a crowdsourcing campaign.
We argue that `pilot study' is a standing term that is being used to refer to small-scale formative studies in crowdsourcing based research.
It is this type of study that we investigated in our paper.
%%% Further, while other terms such as `preliminary study' may be in use,
    % However, the keyword  ``pilot'' was chosen on purpose.
    % A pre-study is a convenient way of identifying needs and defining requirements. For this reason, preliminary studies are often dedicated more space in articles, and it is not uncommon to find articles that dedicate an entire section to a pre-study.
    % We decided to focus on pilot studies as these studies are often under-reported in the literature.

%%% FALSE POSITIVES
% If it was not clear to us whether the pilot engaged the crowd or other participant sample, we included the study in our review.
% This likely results in studies being included even if they did not use crowdsourcing in the pilot study.
% However, our proposed guidelines are nevertheless valid for this set of studies, since they were not clear.

We acknowledge that there may be more reasons that prevent authors from reporting crowd pilot studies more elaborately, which did not surface in our investigation. We therefore cannot treat this work as an exhaustive account of why or how crowd pilot studies are being reported.

% - Missing critical discussion around limitations, including the bias of the selected corpus or the
% chosen keywords, the quantitative approach used for synthesizing papers (i.e., majority of analysis
% is about counting numbers of papers across different categories), etc. 
% But, the current description is more about “their personal opinion/reason” about why they made such a decision
% rather than a critical reflection on (1) why their approach (incl. collected data, qualitative/quantitative analysis) is valid, (2) what other important considerations are missing and
% (3) potential limitations of the chosen approaches.

% ====================
\section{Conclusion}
\label{sec:conclusion}
% ====================
%

In this paper,
we provided an extensive investigation into the state of pilot study reporting in crowdsourcing research. Our systematic screening of over 500 publications at the intersection of HCI and crowdsourcing literature resulted in a corpus of \NUMTOTALARTICLES~articles which we analyzed in depth.
Our analysis revealed that authors are often vague about the extent and content of their crowd pilot studies. Insufficient details pertaining to such pilot studies can hinder replication and reproducibility, and stall the progress of scientific research. We explored the various reasons that drive authors to carry out crowd pilot studies (RQ1), how they are typically reported (RQ2), and what such reports contain (RQ3). Through synthesizing related literature and via a survey study with crowdsourcing researchers in academia and industry, we explored the desirable attributes of a crowd pilot study (RQ4), and the factors that influence the reporting of crowd pilot studies (RQ5). Finally, we explored platform-specific features that can support and facilitate crowd pilot studies (RQ6). Based on our findings, we reflected on how detailed reporting of crowd pilot studies can further aid fair, responsible, and reproducible crowdsourcing research. We presented insights into the trade-offs that authors make while reporting crowd pilot studies and proposed guidelines for reporting them. 
Our proposed guidelines for reporting crowd pilot studies and the APA-inspired way of doing so --- concisely but effectively --- can have important implications on the proliferation of crowdsourcing research, crowdsourcing as a sound scientific method,  and on the anonymous crowd workers who undoubtedly play the most pivotal role in sustaining the crowdsourcing paradigm.

%%%%%%%%%%%%%%%%%%%%%%%%%%%%%%%%%%%%%%%%%%
%%%%%%%%%%%%%%%%%%%%%%%%%%%%%%%%%%%%%%%%%%

% \begin{table}
%   \caption{Frequency of Special Characters}
%   \label{tab:freq}
%   \begin{tabular}{ccl}
%     \toprule
%     Non-English or Math&Frequency&Comments\\
%     \midrule
%     \O & 1 in 1,000& For Swedish names\\
%     $\pi$ & 1 in 5& Common in math\\
%     \$ & 4 in 5 & Used in business\\
%     $\Psi^2_1$ & 1 in 40,000& Unexplained usage\\
%   \bottomrule
% \end{tabular}
% \end{table}

% \begin{table*}
%   \caption{Some Typical Commands}
%   \label{tab:commands}
%   \begin{tabular}{ccl}
%     \toprule
%     Command &A Number & Comments\\
%     \midrule
%     \texttt{{\char'134}author} & 100& Author \\
%     \texttt{{\char'134}table}& 300 & For tables\\
%     \texttt{{\char'134}table*}& 400& For wider tables\\
%     \bottomrule
%   \end{tabular}
% \end{table*}

% \begin{figure}[h]
%   \centering
%   \includegraphics[width=\linewidth]{sample-franklin}
%   \caption{1907 Franklin Model D roadster. Photograph by Harris \&
%     Ewing, Inc. [Public domain], via Wikimedia
%     Commons. (\url{https://goo.gl/VLCRBB}).}
%   \Description{A woman and a girl in white dresses sit in an open car.}
% \end{figure}

  %   \caption{figure caption}
  %   \Description{figure description}

% \begin{acks}
% \end{acks}

\bibliographystyle{ACM-Reference-Format}
\bibliography{cscw24}

% \appendix

\end{document}